\documentclass[prx,twocolumn,english,superscriptaddress,floatfix,longbibliography]{revtex4-2}
\usepackage[utf8]{inputenc}
\usepackage[english]{babel}
\usepackage[T1]{fontenc}
\usepackage{parskip}

\newcommand{\vx}[0]{\mathbf{x}}
\newcommand{\diag}[0]{\text{diag}}

\newcommand{\rsubset}{\rotatebox[origin=c]{90}{$\subset$}}
\usepackage[utf8]{inputenc}
\usepackage{lipsum}

\usepackage{amsmath, amssymb, amsfonts, amsthm, mathtools}
\usepackage{xspace, graphicx, relsize, bm, xcolor}

\usepackage{hyperref}
\usepackage{multirow}
\usepackage{physics}

\usepackage{algorithm}
\usepackage{algpseudocode}
\usepackage{enumitem}

\usepackage{hyperref}

\makeatletter

\renewenvironment{widetext@grid}{%
  \par\ignorespaces
  \setbox\widetext@top\vbox{%
   \vskip15\p@
   \hb@xt@\hsize{%
    \leaders\hrule\hfil
    \vrule\@height6\p@
   }%
   \vskip6\p@
  }%
  \setbox\widetext@bot\hb@xt@\hsize{%
    \vrule\@depth6\p@
    \leaders\hrule\hfil
  }%
  \onecolumngrid
  \let\set@footnotewidth\set@footnotewidth@ii
}{%
  \par
  \twocolumngrid\global\@ignoretrue
  \@endpetrue
}%

\makeatother

\usepackage{sepfootnotes}
\newendnotes{x}
\renewcommand\xnotesize\normalsize

\usepackage{url}
\usepackage{subcaption}
\usepackage{mleftright}

\usepackage{cleveref}

\begin{document}

\title{Quantum kernels through the lens of entangled tensor kernels}

\author{Seongwook Shin}
\email{seongwook_shin@sejong.ac.kr}
\affiliation{Department of Quantum Information Science and Engineering, Sejong University, Seoul 05006, Republic of Korea}
\affiliation{Dahlem Center for Complex Quantum Systems, Freie Universit\"at Berlin, 14195 Berlin, Germany}

\author{Ryan Sweke}
\email{rsweke@aims.ac.za}
\affiliation{African Institute for Mathematical Sciences, Cape Town 7945, South Africa}
\affiliation{Department of Mathematical Sciences, Stellenbosch University, Stellenbosch 7600, South Africa}
\affiliation{National Institute for Theoretical and Computational Sciences (NITheCS), Stellenbosch 7602, South Africa}

\author{Hyunseok Jeong}
\email{h.jeong37@gmail.com}
\affiliation{NextQuantum Innovation Research Center, Department of Physics and Astronomy, Seoul National University, Seoul 08826, Republic of Korea}

\received{29 April 2025}
\accepted[accepted ]{25 April 2026}
\published[published ]{19 May 2026}

\begin{abstract}
Quantum kernel methods are one of the most explored approaches to quantum machine learning. However, the structural properties and inductive bias of quantum kernels are not fully understood. In this work, we introduce the notion of entangled tensor kernels -- a generalization of product kernels from classical kernel theory -- and show that all embedding quantum kernels can be understood as an entangled tensor kernel. We discuss how this perspective allows one to gain insights into both the unique inductive bias of quantum kernels and potential methods for their dequantization. 
\end{abstract}
\maketitle

\section{Introduction}\label{sec:intro}

It is currently an open question to understand the extent to which quantum computers can offer meaningful advantages for practically relevant classical machine learning problems. To this end, a significant amount of effort has been invested in better understanding both the potential and limitations of a wide variety of approaches to quantum machine learning~\mbox{\cite{schuld2018supervised,dunjko2018machine,cerezo2021variational,benedetti2019parameterized}}. One such approach is that of \textit{quantum kernel methods}, which are simply standard kernel-based learning algorithms~\cite{scholkopf2002learning,SVMbook}, in which the classical kernel is replaced with a quantum kernel~\cite{mengoni2019kernel}. 

The first proposals for quantum kernels~\cite{PhysRevLett.122.040504,Havlicek2019} were for what is now referred to as either \textit{embedding quantum kernels}~\cite{Hubregtsen_2022,Gil_Fuster_2024}, or \textit{fidelity quantum kernels}~\cite{gan2023unified}. For these quantum kernels, the high-level idea is as follows: First,  use some data-dependent quantum circuit to map the classical data to quantum feature states (i.e. ``embed'' the classical data into Hilbert space). Then, use a quantum device to measure the fidelity between the quantum feature states. In light of these proposals, the natural question is: Given a specific problem, how should one choose an appropriate quantum kernel? Ideally, one would like a kernel which is efficient to evaluate with a quantum device, hard to evaluate with a classical device, and guaranteed to generalize well for the problem of interest after training on a polynomially sized dataset (i.e. has a suitable \textit{inductive bias}).

Towards this end, one can show that for specially-designed artificial problems, it is possible to construct embedding quantum kernels which yield a rigorous advantage over classical learning algorithms, under cryptographic assumptions~\cite{liu2021rigorous}. However, these results give little insight into how one should choose a quantum kernel for more practically relevant problems, and we are now aware of a variety of potential obstacles in this regard. In particular, one can in fact show that, without taking care, the exponential size of Hilbert space can be a distinct \textit{disadvantage} for quantum kernels. More specifically, it can lead to both poor generalization~\cite{Kubler2021inductive,huang2021power}, and the requirement of exponentially many measurements for evaluating the kernel~\cite{thanasilp2024exponential}. 

Given these obstacles, a variety of approaches have been suggested. One such approach is to introduce a \textit{bandwidth} parameter, which can be shown to facilitate generalization of embedding quantum kernels if chosen appropriately~\cite{Shaydulin_2022, canatar2023bandwidth}. Another approach is to use \textit{trainable} embedding quantum kernels~\cite{Hubregtsen_2022,Rodriguez_Grasa_2025}, which are optimized to match the problem at hand. Finally, \textit{projected quantum kernels} have been suggested~\cite{huang2021power}, in which local quantum feature (reduced) states are used, as opposed to the global fidelity. 

With the goal of further enriching our evolving understanding of quantum kernels, in this work we introduce the notion of an \textit{entangled tensor kernel} -- a generalization of classical product kernels -- and show that all embedding quantum kernels are entangled tensor kernels. We then show how this novel characterization of quantum kernels can be used as lens for their analysis, and in particular to gain a variety of insights into both the potential advantages of quantum kernels, and methods for their dequantization. We provide a more detailed sketch of our contributions in Section~\ref{ss:intro_contribution} below. While this work certainly does not solve the problem of quantum kernel selection, or provide any definitive answers to the question of whether or not one can obtain meaningful advantages with quantum kernels, we hope that the framework of entangled tensor kernels can be a useful tool for the further construction and analysis of both classical and quantum kernels.

\subsection{Contributions}\label{ss:intro_contribution}

\textbf{Entangled tensor kernels:} Our first contribution is to define the notion of an \textit{entangled tensor kernel} (ETK), a natural generalization of classical product kernels. More specifically, any entangled tensor kernel is specified by \textit{local} feature maps (typically into low-dimensional feature spaces), and a core tensor that ``entangles'' the local features.  We analyze the complexity of evaluating ETKs, and provide sufficient conditions for their efficient classical evaluation.
Specifically, whenever the local feature maps can be efficiently evaluated, and the core tensor has a (known) representation as a polynomial bond-dimension matrix product operator (MPO), the ETK can be evaluated efficiently classically. While our primary motivation for introducing ETKs is to provide a novel tool for the analysis of quantum kernels, they are potentially interesting in their own right, as a tensor network based approach to constructing novel classical kernels. As such, ETK's complement the growing body of work aimed at both developing tensor network based machine learning methods, and using tensor networks to gain insight into existing machine learning approaches~\cite{cichocki2016tensor,sengupta2022tensornetworksmachinelearning,anandkumar2014tensor,glasser2019expressive,Sweke2023potential}.

\textbf{Quantum kernels are entangled tensor kernels:} Having defined the notion of an entangled tensor kernel, the main result of this work is the observation that all \textit{embedding} quantum kernels are entangled tensor kernels, whose local feature maps depend only on the data-encoding strategy, and whose core tensor depends only on the data-independent quantum gates, in a well defined way. As such, we show both that entangled tensor kernels provide a novel lens for the analysis and understanding of quantum kernels, and that quantum kernels provide a means for the implementation of ETKs.

\textbf{Inductive bias of quantum kernels through the ETK lens:} Two necessary conditions for achieving a meaningful quantum advantage with a quantum kernel are that it should not be efficiently classically simulable, and when using the quantum kernel one should be able to obtain good generalization for the problem of interest from polynomially many samples -- i.e. the quantum kernel should have the right \textit{inductive bias}. Equipped with the characterization of quantum kernels as ETKs, and our understanding of the complexity of evaluating ETKs, we can immediately see that a necessary (but not sufficient) condition for the hardness of simulating a quantum kernel is that the core tensor of its ETK representation is of super-polynomial bond-dimension. As such, we can identify efficient quantum kernels which generate ETKs whose core tensor is of superpolynomial bond-dimension, as a natural class of quantum kernels for further study. Additionally, the inductive bias of a kernel is determined by its \textit{Mercer decomposition}, and we show how the ETK representation of quantum kernels can facilitate the derivation of their Mercer decomposition, and subsequent analysis of their inductive bias and generalization performance. 

\textbf{Dequantization through the ETK lens:} The fact that all quantum kernels admit an ETK representation, naturally suggests ETKs with polynomial bond-dimension MPO core tensors -- i.e. those which can be evaluated efficiently classically -- as candidate \textit{classical surrogates} for quantum kernels. We develop this idea, and discuss both the potential and limitations of ETKs as classical surrogates for the dequantization of quantum kernels.

\subsection{Structure of this paper}

We begin in Section~\ref{sec:preliminaries} by providing the preliminary material necessary to understand the results of this work. Having done this, we proceed in Section~\ref{sec:ETKs} to define the notion of an entangled tensor kernel. With this in hand, we then show in Section~\ref{Sec:quantumkernelsareETK} that all quantum kernels are entangled tensor kernels. In Section~\ref{sec:landscape} we then discuss the landscape of entangled tensor kernels. More specifically, we use this to discuss when a quantum kernel may provide a potential advantage, and when ETKs may be used as classical surrogates for the dequantization of a quanutm kernel. To further showcase the utility of the ETK lens for the analysis of quantum kernels, we then provide in Section~\ref{ss:example_onelayer} an example of a family of quantum kernels for which we are able to use the ETK representation to obtain a Mercer decomposition. This decomposition then allows one to make concrete statements concerning the inductive bias of these kernels, their generalization capacity, and sample complexity for learning. Finally, we conclude with a summary and a sketch of interesting directions in Section~\ref{sec:conclusion}.

\section{Preliminaries}\label{sec:preliminaries}

\subsection{Notation}\label{ss:notation}

Given a Hilbert space $\mathcal{F}$, we use the notation $\langle \cdot|\cdot\rangle_\mathcal{F}$ for the inner product associated with $\mathcal{F}$. We use both bold font and bra-ket notation to represent vectors in $\mathbb{C}^d$, and we denote the standard dot product in $\mathbb{C}^d$ simply via $\langle \cdot|\cdot\rangle$. For any function $F:\mathbb{C}^{d'}\mapsto\mathbb{C}^{d}$ we represent the $k$'th component of the vector $\ket{F(\vx)}\in \mathbb{C}^d$ as $F_k(\vx)$. In other words, $\ket{F(\vx)} = [F_1(\vx)~F_2(\vx)~\ldots~F_d(\vx)]^\top$.  Additionally, we use the notation $L^2(\mathcal{X},\mu)$ to denote the set of square integrable functions over $\mathcal{X}$ with respect to measure $\mu$. We use $\langle \cdot, \cdot \rangle$ to denote the associated inner product on $L^2(\mathcal{X},\mu)$  -- i.e. for any $f,g\in L^2(\mathcal{X},\mu)$ we have $\langle f, g \rangle = \int\Bar{f}(\vx)g(\vx) \mathrm{d}\mu(\vx)$. 

Additionally, we note that we will make extensive use of tensor network notation. Given the wealth of excellent resources on tensor networks, we do not provide an introduction to them here, but refer the reader to Refs.~\cite{Schollwock_2011,Bridgeman_2017}.

\subsection{Supervised learning}\label{ss:supervised_learning}

In this work we are concerned with algorithms for \textit{supervised learning}. To formalize such algorithms, we start by letting $\mathcal{X}$ denote the set of all possible data points, and $\mathcal{Y}$ a set of all possible labels. We assume the existence of a probability distribution $P$ over $\mathcal{X}\times\mathcal{Y}$, which defines the learning problem, and from which data is drawn. For this work, we will always take $\mathcal{X}=\mathbb{R}^d$, and use $d$ to represent the ``size'' of the learning problem. When $\mathcal{Y}$ is a discrete set (for example $\{0,1\}$) we call the associated learning problem a classification problem, and when $\mathcal{Y}$ is continuous (for example $\mathbb{R}$) we refer to the problem as a regression problem. We will denote the marginal distribution of $P$ on $\mathcal{X}$ with $P_\mathcal{X}$. Additionally, we associate to the learning algorithm a set of functions $\mathcal{H}\subseteq \mathcal{Y}^{\mathcal{X}}$ which we call hypothesis class. Given access to some finite dataset $S = \{(\vx_i,y_i)\sim P\}|_{i = 1}^n$ the goal of the learning algorithm for the supervised learning problem specified by $P$ is to identify the optimal hypothesis $f^*\in\mathcal{H}$, i.e.,  the hypothesis which minimizes the \textit{true risk}, defined via
\begin{equation}
R(f) \coloneqq \mathop{\mathbb{E}}_{(\vx,y)\sim P}\left[\mathcal{L}(y,f(\vx))\right],
\end{equation}
where $\mathcal{L}\colon\mathcal{Y}\times\mathcal{Y}\rightarrow\mathbb{R}$ is some loss function. We note that the true risk cannot be calculated exactly from any finite set of samples without knowledge of $P$ and so we also define the \textit{empirical risk} with respect to the dataset $S$ as 
\begin{equation}
\hat{R}_S(f) \coloneqq \frac{1}{n}\sum_{i = 1}^n\mathcal{L}(y_i,f(\vx_i)).
\end{equation}
As the empirical risk can be calculated from available data, most supervised learning algorithms proceed by minimizing the empirical risk, plus a regularization term which is added to prevent over-fitting to the training data $S$. Moreover, given a learning algorithm with an associated hypothesis class $\mathcal{H}$, one would ideally like to obtain a \textit{generalization bound}, which allows one to put an upper bound on the difference between the true risk and the empirical risk $|\hat{R}_S(f) - R(f)|$ in terms of properties of the hypothesis class, algorithm, dataset $S$ and problem $\mathcal{P}$. 

\subsection{Kernel methods for supervised learning}\label{ss:kernel_methods}

\noindent Some of the simplest learning algorithms possible are those which use \textit{linear functions} as a hypothesis class. In the case of regression, this is the set of functions
\begin{equation}
\mathcal{H}_{\mathrm{lin}} = \{f_{\mathbf{c}}(\vx) = \langle \mathbf{c}|\vx\rangle \,|\, \mathbf{c}\in\mathbb{R}^d\}.
\end{equation}
However, for many problems linear functions are not sufficient to obtain a suitably good solution, and a natural next step is to consider linear functions \textit{in a feature space} $\mathcal{F}$. More specifically, one starts by defining a feature map $F:\mathcal{X}\rightarrow\mathcal{F}$, and then considering the class of functions
\begin{equation}
\mathcal{H}_{F} = \{f_{\mathbf{c},F}(\vx) = \langle \mathbf{c}|F(\vx)\rangle_\mathcal{F} \,|\, \mathbf{c}\in\mathcal{F}\}.
\end{equation}
For any such feature map, one can define an associated \textit{kernel} $K_F:\mathcal{X}\times\mathcal{X}\mapsto\mathbb{R}$ via
\begin{equation}
K_F(\mathbf{x},\mathbf{x}') = \langle F(\mathbf{x})|F(\mathbf{x}')\rangle_\mathcal{F}.
\end{equation}
We note that for real (complex) kernels the feature space $\mathcal{F}$ can be any real (complex) Hilbert space, even infinite dimensional. However, whenever $\mathcal{F}$ is a finite $d$-dimensional real Hilbert space, there exists an equivalent feature map $\tilde{F}:\mathcal{X}\mapsto\mathbb{R}^d$ satisfying
\begin{equation}\label{eq:equiv}
\langle F(\cdot)|F(\cdot)\rangle_\mathcal{F} = K_F(\cdot,\cdot) = K_{\tilde{F}}(\cdot,\cdot) = \langle \tilde{F}(\cdot)|\tilde{F}(\cdot)\rangle,
\end{equation}
where $\langle \cdot | \cdot\rangle$ is the standard dot product in $\mathbb{R}^d$. To construct $\tilde{F}$, let $\{v_i\}$ be an orthonormal basis for $\mathcal{F}$, and define $\varphi:\mathcal{F}\rightarrow \mathbb{R}^{d}$ via $\varphi(\sum_i a_iv_i) = [a_1 \ldots a_d]^\top$. Then, $\tilde{F} = \varphi\circ F$ can be seen to satisfy Eq.~\eqref{eq:equiv}. We will use this observation in Section~\ref{sec:ETKs} to restrict ourselves to considering feature maps to $\mathbb{R}^d$ for finite $d$.

In many settings, access to the kernel function $K_F$ is sufficient to identify the optimal hypothesis in $\mathcal{H}_{F}$, with respect to a suitably regularized empirical risk. In this work, we call learning algorithms which only use access to the kernel function \textit{kernel methods}. While we refer to Refs~\cite{scholkopf2002learning, SVMbook} for more detailed information on such algorithms, which includes standard methods such as support vector machines and kernel ridge regression, we mention two important facts here:
\begin{enumerate}
\item For certain feature maps, evaluating the kernel function $K_F$ can be done much more efficiently than by explicitly constructing the vectors $F(\mathbf{x})$ and $F(\mathbf{x}')$ then taking the inner product. Using feature maps which allow one to exploit this, is called \textit{the kernel trick}.
\item It is in fact not even necessary to start by specifying a feature map. Indeed,  for \textit{any} symmetric positive semidefinite function $K:\mathcal{X}\times\mathcal{X}\mapsto\mathbb{R}$, there is some feature map for which $K$ is the kernel. As such, in practice, one typically simply chooses a suitable kernel, as opposed to a feature map.
\end{enumerate}

\subsubsection{Inductive bias of kernel methods}\label{sss:inductive_bias}

The first step in any kernel method is choosing a kernel function, and as such one would ideally like to understand the \textit{inductive bias} associated with a specific kernel -- or in other words, the set of problems for which a specific kernel is well suited. To this end, it is useful to analyze the \textit{kernel integral operator} associated with a kernel. To be more specific, for any non-negative measure $P_\mathcal{X}$ on $\mathcal{X}$, we define the kernel integral operator $T_{(K,P_\mathcal{X})}$ via
\begin{equation}
    T_{(K,P_\mathcal{X})}(f)(\vx) := \int_{\mathcal{X}}K(\vx,\mathbf{y})f(\mathbf{y})\mathrm{d}P_{\mathcal{X}}(\mathbf{y}).
\end{equation}
It then follows from \textit{Mercer's theorem} that there exists a (possibly infinite) spectral decomposition of $T_{(K,P_\mathcal{X})}$, with eigenvalues $\{\gamma_i\}$ and associated eigenfunctions $\{e_i\}$, which satisfies:

\begin{enumerate}
\item All eigenvalues are non-negative -- i.e. $\gamma_i\geq 0$ for all $i$.
\item The eigenfunctions $\{e_i\}$ form an orthonormal basis for $L^2(\mathcal{X},P_\mathcal{X})$.
\item The kernel $K$ admits a \textit{Mercer decomposition}
\begin{align}
K(\vx,\vx') &= \sum_{i}^{d}\gamma_i e_i(\vx)e_i(\vx')\\
&=\bra{e(\vx)}\Gamma\ket{e(\vx)},\label{eq:mercer_form}
\end{align}
where $d$ is the number of non-zero eigenvalues, $|e(\mathbf{x})\rangle = |e_1(\mathbf{x})\ldots e_d(\mathbf{x})\rangle$ and $\Gamma$ is the diagonal matrix with $\Gamma_{i,i} = \gamma_i$.
\end{enumerate}

For certain kernel methods, such as kernel ridge regression (KRR), the inductive bias of the method when using kernel $K$ is closely related to the the spectral decomposition of $T_{(K,P_\mathcal{X})}$ (which for convenience we sometimes refer to as the Mercer decomposition of $K$). More specifically, as $\{e_i\}$ is a basis for $L^2(\mathcal{X},P_\mathcal{X})$, we can decompose any $f\in L^2(\mathcal{X},P_\mathcal{X})$ as $f = \sum_{i}a_i e_i$. At a high level,  then is biased towards functions which are supported on eigenfunctions $e_i$ corresponding to large eigenvalues $\gamma_i$. Said another way, KRR is biased towards functions whose components $\{a_i\}$ are ``well-aligned''  with the spectrum of $T_{(K,P_\mathcal{X})}$. 

These high level statements can be made more precise through notions such as \textit{kernel-target alignment}~\cite{cristianini2001kernel} and \textit{task-model alignment}~\cite{Canatar2021spectral,canatar2023bandwidth}, both of which provide different measures of the alignment between the spectrum of the kernel integral operator and the principal components of the target function, when written in the eigenbasis of $T_{(K,P_\mathcal{X})}$. The latter in particular is useful, as one can prove generalization bounds in terms of the task-model alignment, which in turn allow one to bound the sample complexity required to learn specific functions in a meaningful sense.

\subsection{Quantum kernel methods}\label{ss:_QKM}

As discussed in Section~\ref{sec:intro}, quantum kernel methods are kernel based learning algorithms in which the kernel function is evaluated using a quantum computational device~\cite{mengoni2019kernel,PhysRevLett.122.040504,Havlicek2019}. In this work we will consider \textit{embedding quantum kernels}~\cite{Gil_Fuster_2024}, which have the form
\begin{equation}\label{eq:quantum_kernel}
    K_U(\vx,\vx') := \abs{\bra{0}^{\otimes n}U^{\dag}(\vx)U(\vx')\ket{0}^{\otimes n}}^2,
\end{equation}
where $U(x)$ is some data-dependent quantum circuit which serves as a feature map, ``embedding'' classical data into $n$-qubit quantum states. To see that this is indeed a valid (real) kernel function, first recall that the set of Hermitian matrices $\mathrm{Herm}(2^n)$, equipped with the Hilbert-Schmidt inner product, defined via $\langle A| B\rangle_{\mathrm{HS}} = \mathrm{Tr}[A^\dagger B]$, is a real Hilbert space. Then, note that if we define the feature map $\mathcal{U}:\mathcal{X}\mapsto\mathrm{Herm}(2^n)$ via
\begin{align}
\mathcal{U}(\vx) &= U(\vx)\ket{0}^{\otimes n}\bra{0}^{\otimes n} U^{\dagger}(\vx)\\ &:=\rho(x)
\end{align}
then have that
 \begin{align}
    K_U(\vx,\vx') &= \langle \mathcal{U}(\vx)|\mathcal{U}(\vx')\rangle_\mathrm{HS},\\
    &= \mathrm{Tr}\left[\rho(\vx)\rho(\vx')\right].
\end{align}
Said another way, $K_U(\vx,\vx')$ is simply the Hilbert-Schmidt inner product of the  (pure) \textit{feature states} $\rho(\vx)$ and $\rho(\vx')$, and therefore a valid real kernel. Additionally, we note that Eq.~\eqref{eq:quantum_kernel} is the \textit{fidelity} of the feature states $\rho(\vx)$ and $\rho(\vx')$, which motivates the name of \textit{fidelity quantum kernels}~\cite{gan2023unified}. 

\begin{figure}
    \centering
    \includegraphics{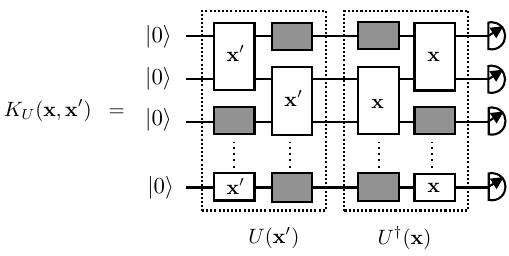}
    \caption{A circuit diagram for quantum kernel evaluation -- i.e. for the evaluation of Eq.~\eqref{eq:quantum_kernel}. White boxes represent data-dependent quantum gates, while gray ones are non-parametrized gates.} 
    \label{fig:QK_evaluation}
\end{figure}

Given this, we see that to specify an embedding quantum kernel, one needs to define a data-dependent quantum circuit $U(x)$. Such circuits will in general consist of both data-dependent gates and fixed non-parameterized gates. There also exist recent proposals for \textit{trainable} quantum kernels~\cite{Hubregtsen_2022,Rodriguez_Grasa_2025}, including gates parameterized by trainable parameters, for which everything that follows would hold once these trainable parameters are fixed. Given such a circuit, the kernel function $K_U(\vx,\vx')$ can then be approximated in multiple ways. One way is via repetition of the quantum circuit illustrated in Fig.~\ref{fig:QK_evaluation}. Specifically, one first implements the circuit $U(\vx)$, then the circuit $U^{\dagger}(\vx')$, followed by a measurement in the computational basis. Another method is to prepare both $\rho(\vx)$ and $\rho(\vx')$, then use a SWAP test to evaluate their inner product. These methods are both efficient whenever $U(\vx)$ can be implemented efficiently, and we refer to Ref.~\cite{shastry2023shotfrugalrobustquantumkernel} for a comparison of their relative advantages and disadvantages.

Finally, we stress that the essential difference between quantum and classical kernel methods is the way in which the kernel function is evaluated. As such, all theory pertaining to classical kernel methods -- i.e. inductive bias, sample complexity, generalization etc -- is also applicable to quantum kernel methods. Indeed, this classical theory has already been used to gain significant insight into the inductive bias of quantum kernels~\cite{Kubler2021inductive,huang2021power,Shaydulin_2022, canatar2023bandwidth}. Additionally, we stress that in this work we are uniquely concerned with \textit{embedding} quantum kernel methods as described above, and do not consider alternative quantum kernel methods such as \textit{projected} quantum kernels~\cite{huang2021power}. For a detailed benchmarking study of both embedding and projected quantum kernels we refer to Ref.~\cite{Schnabel_2025}.

\section{Entangled tensor kernels}\label{sec:ETKs}

There are a variety of situations in which one might want to build new kernels out of existing kernels. In this work, we consider the following two scenarios:

\textbf{Extending the data domain:} Imagine one has two kernels $K_1: \mathcal{X}_1\times \mathcal{X}_1 \mapsto \mathbb{R}$ and ${K_2 : \mathcal{X}_2\times \mathcal{X}_2 \mapsto\mathbb{R}}$, for different data domains $\mathcal{X}_1\neq\mathcal{X}_2$. Sometimes, one would like a kernel which can be used for learning on the product domain $\mathcal{X} = \mathcal{X}_1\times\mathcal{X}_2$. This would be the case when a dataset is enriched by the addition of new distinct features, either through additional data collection or the availability of new data (for example, when a new version of a census contains new categories).

\textbf{Enhancing the feature space:} Imagine one has two different kernels $K_1: \mathcal{X}\times \mathcal{X} \mapsto \mathbb{R}$ and ${K_2 : \mathcal{X}\times \mathcal{X} \mapsto \mathbb{R}}$, for the \textit{same} data domain $\mathcal{X}$. Let's denote the feature map associated with $K_i$ via $F^{(i)}:\mathcal{X}\mapsto\mathcal{F}_i$. Sometimes, the feature spaces $\mathcal{F}_i$ may not be ``expressive enough'' to provide good solutions, and one would like to build a new kernel $K:\mathcal{X}\times\mathcal{X}$, using a feature map with the higher dimensional feature space $\mathcal{F}_1\otimes\mathcal{F}_2$.

In what follows, we will introduce the notion of \textit{entangled tensor kernels}, which can be used in both of the above situations, in the case when one knows and can evaluate the feature maps of the constituent kernels used as building blocks. In order to gain some intuition, we start with the simple existing method of \textit{product kernels}~\cite{Wainwright2019}.

\subsection{Warm up: Product kernels}\label{ss:prod_kernels}

Let's assume that we are in the first situation described above. More specifically, let's assume that we are given two kernels $K_1: \mathcal{X}_1\times \mathcal{X}_1 \rightarrow \mathbb{R}$ and $K_2 : \mathcal{X}_2\times \mathcal{X}_2 \rightarrow \mathbb{R}$, and we would like to use these to construct a new kernel $K:(\mathcal{X}_1 \times \mathcal{X}_2)\times(\mathcal{X}_1 \times \mathcal{X}_2)\rightarrow\mathbb{R}$ for the domain $\mathcal{X}=\mathcal{X}_1\times\mathcal{X}_2$. One way to do this is to define
\begin{equation}\label{eq:tpk}
    K((\vx_1,\vx_2),(\vx'_1,\vx'_2)) := K_1(\vx_1,\vx'_1)K_2(\vx_2,\vx'_2).
\end{equation}
To see that this is a valid kernel, and to provide a foundation for the entangled tensor kernels we introduce in Section~\ref{ss:ETK}, it is instructive to think about this kernel from the feature map perspective. To this end let $F^{(i)}:\mathcal{X}_i\mapsto \mathcal{F}_i$ be a feature map for $K_i$. We can then easily see that $F:\mathcal{X}_1 \times \mathcal{X}_2\rightarrow \mathcal{F}=\mathcal{F}_1\otimes\mathcal{F}_2$ defined as
\begin{equation}\label{eq:tensor_product_feature}
|F(\vx_1,\vx_2)\rangle = | F^{(1)}(\vx_1)\rangle\otimes | F^{(2)}(\vx_2)\rangle,
\end{equation}
satisfies
\begin{widetext}
\begin{align}
\langle F(\vx_1,\vx_2)|F(\vx'_1,\vx'_2)\rangle_\mathcal{F} 
&= \left(\langle F^{(1)}(\vx_1)|\otimes  \langle F^{(2)}(\vx_2)|\right) \left( | F^{(1)}(\vx'_1)\rangle\otimes | F^{2}(\vx'_2)\rangle\right)\\
&= \langle F^{(1)}(\vx_1) | F^{(1)}(\vx_1')\rangle_{\mathcal{F}_1}\langle F^{(2)}(\vx_2) | F^{(2)}(\vx_2')\rangle_{\mathcal{F}_2}\\
&=K_1(\vx_1,\vx'_1)K_2(\vx_2,\vx'_2)\\
&= K((\vx_1,\vx_2),(\vx'_1,\vx'_2)).
\end{align}
In particular, we see from Eq.~\eqref{eq:tensor_product_feature}, that for any kernel defined via Eq.~\eqref{eq:tpk}, the feature vector $|F(\vx_1,\vx_2)\rangle$ is simply the tensor product of feature vectors $|F^{(1)}(\vx_1)\rangle$ and $|F^{(2)}(\vx_2)\rangle$. This motivates the name \textit{product kernels}~\cite{Wainwright2019} for such kernels. We note that the construction above clearly generalizes to $N$ kernels $\{K_i\,|\,i\in [N]\}$.
Additionally, we also note that the Mercer decomposition of this new product kernel can be obtained easily from the Mercer decompositions of the constituent kernels. In particular, one has 
\begin{align}
    K((\vx_1,\vx_2),(\vx'_1,\vx'_2)) =\sum^{d_1}_{i}\sum_j^{d_2}\gamma^{(1)}_i\gamma^{(2)}_je_i^{(1)}(\vx_1)e_j^{(2)}(\vx_2)e_i^{(1)}(\vx'_1)e_j^{(2)}(\vx'_2)
\end{align} 
where $\{\gamma^{(1,2)}_i\}$ and $ \{e^{(1,2)}_i\}$ are the non-zero eigenvalues and corresponding eigenfunctions for the respective kernels, and $d_{1,2}$ is used to denote the number of non-zero eigenvalues. In particular, one can notice that here the eigenfunctions and eigenvalues of the product kernel are simply the products of those of the individual kernels.
\end{widetext}

Let us now move on to the second situation. In particular, let's assume that we are given two kernels $K_1: \mathcal{X}\times \mathcal{X} \mapsto \mathbb{R}$ and ${K_2 : \mathcal{X}\times \mathcal{X} \mapsto \mathbb{R}}$, and we would like to build a new kernel $K:\mathcal{X}\times\mathcal{X}\mapsto\mathbb{R}$, with a different feature space. Similarly to before, one option is to simply take the product of the kernels, i.e. 
\begin{equation}
K(\vx,\vx') := K_1(\vx,\vx')K_2(\vx,\vx').
\end{equation}
Once again, from the explicit feature map perspective, one has that
\begin{equation}
K(\vx,\vx') = \langle F(\vx)|F(\vx')\rangle_\mathcal{F}, 
\end{equation}
where $F:\mathcal{X}\rightarrow \mathcal{F}=\mathcal{F}_1\otimes\mathcal{F}_2$ via
\begin{equation}
|F(\vx)\rangle=  |F^{(1)}(\vx)\rangle\otimes | F^{(2)}(\vx)\rangle.
\end{equation}
In particular, we see that this new kernel has the enhanced feature space $\mathcal{F}=\mathcal{F}_1\otimes\mathcal{F}_2$. Again, we note that this construction again clearly generalizes to the case of $N$ kernels $\{K_i\}$, with feature maps $\{F^{(i)}\}$. We stress however that in this case the Mercer decomposition of the product kernel is not immediately obtained from the Mercer decompositions of the constituent kernels, as the functions $\{e_i(\vx)e_j(\vx)\}|_{i,j}$ are not necessarily orthogonal themselves.

In summary, we see that taking the product of two different kernels -- or alternatively, the tensor product of feature vectors -- allows one to obtain a new kernel with either:
\begin{enumerate}
\item An extended domain and feature space (if the constituent kernels act on different domains).
\item An extended feature space (if the constituent kernels act on the same domain).
\end{enumerate}

\subsubsection{Computational complexity of evaluating product kernels}\label{ss:complexity_product}

Consider a kernel $K_F$, for which $K_F(\cdot,\cdot) = \langle F(\cdot)|F(\cdot)\rangle$, where $F:\mathcal{X}\mapsto\mathbb{R}^d$. Assuming knowledge of the feature map $F$, the naive way to evaluate $K_F(\vx,\vx')$ is to first construct the feature vectors $|F(\vx)\rangle, |F(\vx')\rangle \in \mathbb{R}^d$, and then take the inner product, which requires~$d$ multiplications. However, as mentioned in Section~\ref{ss:kernel_methods}, for certain kernels $K_F(\vx,\vx')$ can in fact be evaluated much more efficiently than via the naive method. Indeed, this \textit{kernel trick} is central to the success of kernel methods, as it facilitates the use of extremely high-dimensional (even infinite) feature spaces~\cite{scholkopf2002learning,SVMbook}. Here, we briefly examine the extent to which a ``kernel trick'' is possible for product kernels.

To this end, lets consider a product kernel built from $N$ kernels $\{K_i:\mathcal{X}\times\mathcal{X}\mapsto\mathbb{R}\}$, where each $K_i$ is associated with a feature map $F^{(i)}:\mathcal{X}\rightarrow\mathbb{R}^{d}$. Additionally, let us assume that each kernel $K_i$ admits a kernel trick which allows the kernel to be evaluated with $T_i < d$ multiplications. We now have the following options for evaluating the product kernel $K(\vx,\vx')$:

\begin{enumerate}
\item First explicitly construct the feature vectors $|F(\vx)\rangle,|F(\vx')\rangle\in\mathbb{R}^{d^N}$, then take the inner product. This requires $O(d^N)$ multiplications -- i.e. the cost scales exponentially with the number of constituent kernels.
\item For each $i \in [N]$, evaluate $K_i(\vx,\vx')$ by first evaluating $|F^{(i)}(\vx)\rangle,|F^{(i)}(\vx')\rangle\in\mathbb{R}^{d}$, then taking the inner product. Finally, take the product of all constituent kernel evaluations. This requires $O(Nd)$ multiplications.
\item For each $i\in[N]$, simply use the appropriate kernel trick to evaluate $K_i(\vx,\vx')$, then take the product of all constituent kernel evaluations. This requires $\prod_i T_i < Nd$ multiplications.
\end{enumerate}
As such, we see that product kernels do indeed admit kernel tricks, which allows for the kernel to be evaluated significantly more efficiently than via the naive method of explicitly constructing the feature vector and taking the inner product.

\subsection{Entangled tensor kernels}\label{ss:ETK}

In this section, we generalize the simple product kernel construction to obtain \textit{entangled tensor kernels} (ETKs). However, for ease of presentation, from this point on we will consider explicitly only the second perspective above -- i.e. constituent kernels which all share the same domain. It should however be clear from the discussion in the previous section how this can be easily generalized to the setting of constituent kernels with different domains. Additionally, in this section we will also only consider feature maps $F:\mathcal{X}\mapsto\mathbb{R}^d$, for some finite $d$. However, as we have discussed in Section~\ref{ss:kernel_methods}, for any kernel $K_F$ defined via a feature map $F:\mathcal{X}\mapsto \mathcal{F}$, where $\mathcal{F}$ is any finite $d$-dimensional real Hilbert space, one can always find an equivalent feature map $\tilde{F}:\mathcal{X}\mapsto\mathbb{R}^d$, for which $K_F = K_{\tilde{F}}$.

With this in mind, lets again consider two kernels $K_1: \mathcal{X}\times \mathcal{X} \rightarrow \mathbb{R}$ and ${K_2 : \mathcal{X}\times \mathcal{X} \rightarrow \mathbb{R}}$, with corresponding feature maps $F^{(1)}:\mathcal{X}\rightarrow \mathbb{R}^{d_1}$ and $F^{(2)}:\mathcal{X}\rightarrow \mathbb{R}^{d_2}$. Given a ${d_1d_2\times d_1d_2}$ positive semidefinite (PSD) matrix $C$, we now define the entangled tensor kernel ${K_C:\mathcal{X}\times\mathcal{X}\rightarrow\mathbb{R}}$ via
\begin{equation}
K_C(\vx,\vx') = \bra{F^{(1)}(\vx)}\bra{F^{(2)}(\vx)}C\ket{F^{(1)}(\vx')}\ket{F^{(2)}(\vx')}.
\end{equation}
To see that this is indeed a valid kernel, first note that we can always write $C = B^\top B$, where $B$ is an $m\times d_1d_2$ matrix, and $m=\mathrm{rank}(C)$. With this, we can see that
\begin{equation}\label{eq:etk_def}
K_C(\vx,\vx') =  \langle F_C (\vx)| F_C (\vx')\rangle,
\end{equation}
where $F_C:\mathcal{X}\mapsto \mathbb{R}^m$ via
\begin{equation}\label{eq:entangling}
| F_C (\vx)\rangle = B\ket{F^{(1)}(\vx)}\ket{F^{(2)}(\vx)}.
\end{equation}
As $K_C$ is calculated via an inner product in a feature space, it is indeed symmetric and positive semidefinite, and therefore a valid kernel. Moreover, from Eq.~\eqref{eq:entangling} we can see that for an ETK $K_C$, the feature vector $|F_C (\vx)\rangle$ is no longer necessarily a tensor product of local feature vectors -- i.e. it may not be possible to write $|F_{C}(\vx)\rangle = \ket{\tilde{F}^{(1)}(\vx)}\ket{\tilde{F}^{(2)}(\vx)}$ for all $\vx\in\mathcal{X}$, for two local feature maps $\tilde{F}^{(1)}:\mathcal{X}\rightarrow \mathbb{R}^{d_1}$ and $\tilde{F}^{(2)}:\mathcal{X}\rightarrow \mathbb{R}^{d_2}$. By analogy with the definition of entanglement for a many-body quantum state, we \textit{define} the feature map $F_C$ as being entangled, whenever it cannot be written as the tensor product of local feature maps. Whenever $F_C$ is an entangled feature map according to this definition, then we call the kernel $K_C$ defined via eq.~\eqref{eq:etk_def} an \textit{entangled tensor kernel}. In particular, we stress that the word ``entangled" is used in this manuscript simply to represent the impossibility of writing a feature map as a tensor product of local feature maps.


Additionally, note that when $C$ is full rank, then $K_C$ is calculated via the inner product in $\mathbb{R}^{d_1d_2}$ defined by $C$. Moreover, when $C = I$ this is the standard dot product, and one recovers the product kernel.  To aid in developing intuition, we have provided in Appendix~\ref{app:examples_of_ETK} a variety of examples, illustrating how various classical kernels can be understood as ETKs.

We stress however that unlike the case of product kernels, the Mercer decomposition of an ETK does not follow immediately from the Mercer decomposition of the constituent kernels. We address this issue in Appendix~\ref{app:mercer_decomp}, where we provide an algorithm for obtaining the Mercer decomposition of an ETK.

Finally, while the above example only considered two constituent kernels, it should be clear that the construction can easily be generalized to multiple constituent kernels. Specifically, given $N$ feature maps $F^{(k)} : \mathcal{X} \rightarrow \mathbb{R}^{d_k}$ (some of which could be the same) we can create a new feature map by entangling the feature vectors of $\{F^{(i)}\}$ via (the square root of) any $\prod_k d_k \times\prod_k d_k$ PSD matrix $C$. More specifically, in this case one has

\begin{equation}\label{eq:etk_def_1}
    K_{C}(\vx,\vx') = \bra{F(\vx)}C\ket{F(\vx')},
\end{equation} 
where we have defined 
\begin{equation}\label{eq:etk_def_2}
    \ket{F(\vx)} := \bigotimes_{k=1}^N\ket{F^{(k)}(\vx)}. 
\end{equation}
In the following sections, it will be extremely convenient to use tensor network notation~\cite{Schollwock_2011,Bridgeman_2017} for the representation of such kernels. To this end, we note that one can represent the ETK $K_C$ using tensor network notation via
\begin{equation}\label{eq:ETK_TN_form}
    K_C(\vx,\vx') = \vcenter{\hbox{\includegraphics[height=0.2\linewidth]{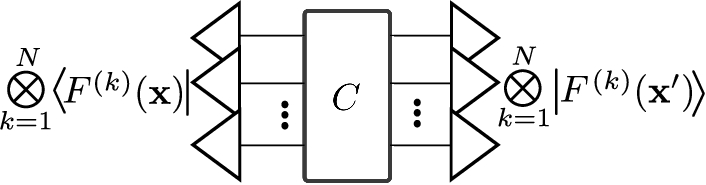}}},
\end{equation}
where each small (left-)right-hand side looking triangle represents the local feature vector $\left(\bra{F^{(k)}(\vx)}\right)$ $\ket{F^{(k)}(\vx)}$ and the line represents the $d_k$-dimensional index.
Alternatively, one could also write
\begin{equation}
K_C(\vx,\vx') =  \langle F_C (\vx)| F_C (\vx')\rangle
\end{equation}
with
\begin{align}
| F_C (\vx)\rangle &= B\ket{F(\vx)},
\end{align}
where $C = B^\top B$. Analogously to the two–feature-map case, we call the feature map \emph{entangled} if $B\ket{F(\vx)}$ does not admit a tensor-product decomposition, or equivalently $B\neq \bigotimes_{k=1}^NB_k$ for any choice of $B_k$s, we call this feature map \textit{entangled.}

\subsection{Computational complexity of evaluating entangled tensor kernels}\label{ss:complexity}

We have seen in Section~\ref{ss:complexity_product} that whenever the constituent kernels of a product kernel admit a kernel trick, then the product kernel itself admits a kernel trick, and can be evaluated efficiently.
In this section, we examine the extent to which a kernel trick is possible for entangled tensor kernels. To this end, similarly to Section~\ref{ss:complexity_product}, lets consider an ETK $K_C:\mathcal{X}\times\mathcal{X}\rightarrow \mathbb{R}$, as in Eqs.~\eqref{eq:etk_def_1} and~\eqref{eq:etk_def_2}, built from $N$ local feature maps $\{F^{(i)}:\mathcal{X}\rightarrow\mathbb{R}^{d}\,|\,i\in[N]\}$ and some arbitrary PSD core matrix $C$. Unfortunately, in this case it is not immediately clear whether one can evaluate $K_C(\vx,\vx')$ more efficiently than by naively constructing $|F(\vx)\rangle,|F(\vx')\rangle\in\mathbb{R}^{d^N}$ and then evaluating $\langle F(\vx)|C|F(\vx')\rangle$, which requires $O(d^{2N})$ multiplications. However, by exploiting the tensor product structure of the feature space, we can indeed sometimes evaluate $K_C$ more efficiently than via this naive method.

To be more specific, as the feature space of the ETK has the tensor product structure $\mathcal{F} = \left(\mathbb{R}^d\right)^{\otimes N}$, one can represent the core tensor as an $N$-site matrix product operator (MPO). In this case, we have that 
\begin{equation}\label{eq:MPO_contract}
K_C(\vx,\vx') = \vcenter{\hbox{\includegraphics[height=0.22\linewidth]{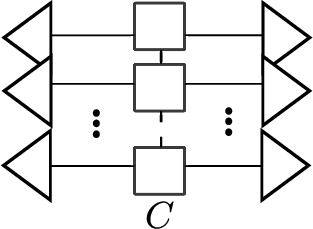}}},
\end{equation}
where for convenience, we have omitted labeling the small triangles (local feature vectors) in the above diagram. Moreover, the tensor network on the right hand side of Eq.~\eqref{eq:MPO_contract} can be contracted using $O(d^2\chi^2N)$ multiplications, where $\chi$ is the \textit{bond-dimension} of the MPO representation of~$C$~\cite{Schollwock_2011,Bridgeman_2017}. In particular, we see that if $\chi = \mathrm{poly}(N)$, then $K_C(\vx,\vx')$ can be evaluated efficiently with respect to $N$. To summarize, if $C$ is specified via a polynomial bond-dimension MPO, then there does indeed exist a kernel trick, via tensor network contractions, which allows one to evaluate $K_C(\vx,\vx')$ significantly more efficiently than via the naive approach. Said another way, given an ETK specified by a core tensor $C$, the existence of a polynomial bond-dimension MPO representation of the core tensor is a \textit{sufficient condition} for the efficient evaluation of the ETK, assuming that one can efficiently evaluate all local feature maps. Additionally, with this perspective we can give an alternative equivalent definition of an \textit{entangled} feature map, which clarifies the sense in which we use the word entangled in this manuscript. More specifically, the kernel defined via Eq.~\eqref{eq:MPO_contract} has an \textit{entangled} feature map whenever the minimum bond dimension of the MPO representation of $C$ is strictly greater than 1. Conversely, if the maximum bond dimension of this MPO is equal to 1, then $K_C$ has a product feature map and is therefore a product kernel. Note that such product kernels are still ETKs in our terminology; they correspond to ETKs whose core tensor C is a product MPO. Therefore, with slight abuse of notation, we also refer to product kernels as ETKs, with a \textit{product} feature map. 

Given this, if one wants to specify an ETK, with the guarantee that it can be evaluated efficiently, it seems natural to simply specify a polynomial bond-dimension MPO core tensor.
Unfortunately however, the situation is not as simple as it might seem. In particular, the matrix defined by the MPO (after appropriate grouping of indices) is required to be positive semidefinite, and unfortunately it is not immediately clear how one can ensure positive semidefiniteness of the global core tensor through constraints on the local tensors which specify the MPO representation. In fact, if one attempts to define an ETK by specifying an MPO core tensor, it is provably $\mathsf{NP}$-hard to check that the given MPO indeed defines a positive semidefinite matrix, and therefore a valid ETK core tensor~\cite{kliesch2014matrix}.

Luckily however, this is exactly the same problem encountered when trying to use matrix product operators to parameterize density matrices of many-body quantum systems, which are also required to be positive semidefinite (and Hermitian)~\cite{kliesch2014matrix}. As such we can adopt the solution from that context, which is to use \textit{locally purified matrix product operators} (LP-MPOs), which are positive semidefinite and Hermitian by construction~\cite{Werner_2016,Cuevas_2013}. More specifically, we can specify the core tensor via an MPO representation $\{A^{1},\ldots,A^{(N)}\}$ for an operator $X$, from which the core tensor is defined via $C = XX^\dagger$, which is guaranteed to be positive semidefinite (and Hermitian). Diagrammatically, such core tensors are represented as follows
\begin{equation}
\vcenter{\hbox{\includegraphics[height=0.25\linewidth]{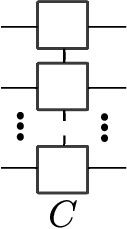}}} = \vcenter{\hbox{\includegraphics[height=0.25\linewidth]{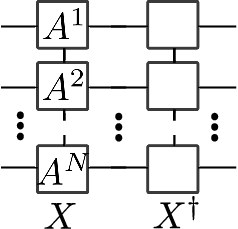}}}.
\end{equation}
We note that each tensor $A^{(i)}$ in the MPO representation of $X$ has dimensions $d\times\chi^2\times p$, where $\chi$ is the bond-dimension and $p$ is the so-called purification dimension. When the core tensor $C$ is represented by a tensor-network in this way, we will refer to it as a locally purified core tensor. From the above discussion, it should now be clear that entangled tensor kernels, specified via locally purified core tensors with polynomial (in $N$) bond-dimension and purification dimension, can be both evaluated efficiently, and efficiently verified as valid kernels. 

\section{Quantum kernels are entangled tensor kernels\label{Sec:quantumkernelsareETK}}

In this section, we provide evidence for the following statement:

\begin{center}
    \textit{All embedding quantum kernels are entangled tensor kernels.}
\end{center}

In particular, we show that \textit{any} embedding quantum kernel $K_U$ can be written as an ETK, whose local feature maps depend only on the data-dependent gates of the data-dependent quantum circuit $U$. As we will discuss at length in Section~\ref{sec:landscape}, this perspective on quantum kernels is interesting, because the ETK form of a quantum kernel allows insights into both its inductive bias and classical simulability, which are not necessarily clear from the circuit description of the quantum kernel.

To show this, we start in Section~\ref{ss:standard_form} by introducing a \textit{standard form} for the data-dependent quantum circuits which define quantum kernels. This standard form then allows us, in Section~\ref{Sec:quatnumkernelsinTN}, to use tensor network techniques to express any embedding quantum kernel as an ETK.

\subsection{A standard form for data-dependent quantum circuits}\label{ss:standard_form}

We start by noting that any data-dependent quantum circuit $U(\vx)$ can be written as a quantum circuit containing only non-parameterized gates, and data-dependent single-qubit gates of the form
\begin{equation}
U_{\phi}(\vx) = e^{-iZ\phi(\vx)/2},
\end{equation}
where $\phi:\mathbb{R}^d\rightarrow\mathbb{R}$ is some pre-processing function, and $Z$ is the single qubit Pauli $Z$ operator. This is achieved via successive Givens rotations~\cite{Juha2004efficient}, as described in detail in Appendix~\ref{app:decomposition_into_standard_form}. Diagrammatically, this decomposition yields
\begin{align}
    U(\vx) &= \vcenter{\hbox{\includegraphics[height=0.3\linewidth]{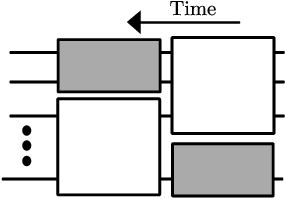}}}\label{eq:original_circuit}\\
    &=\vcenter{\hbox{\includegraphics[height=0.3\linewidth]{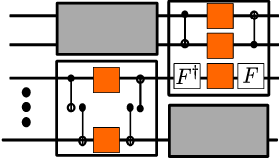}}}\label{eq:first_standard_circuit}
\end{align} 
where gray boxes consist of non-parametrized unitaries, white boxes are arbitrary data-dependent unitaries, orange boxes are data-dependent Pauli-Z rotations, and the $F$ gates are defined in Appendix~\ref{app:decomposition_into_standard_form}. With this in hand, we will from now on assume that any data-dependent quantum circuit is in this form.

In addition, it will be convenient to further rewrite the circuit $U(\vx)$ as alternating layers of data-independent and data-dependent gates. Diagramatically, we have
\begin{align}
    U(\vx) &=\vcenter{\hbox{\includegraphics[height=0.3\linewidth]{Eqs/Eq_unitary_2.eps}}}\label{eq:first_standard_circuit}\\
     &=\vcenter{\hbox{\includegraphics[height=0.32\linewidth]{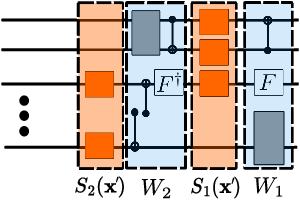}}}\label{eq:circuit_standard_form}\\
    &=\prod_{j=1}^L S_j(\vx)W_j,\label{eq:standard_expression}
\end{align} 
where $\{W_j\}$are non-parametrized (data-independent) unitaries, and we define
$S_j(\vx) :=\left(\bigotimes_{k=1}^{n}e^{-i\phi_{jk}(\vx)Z_{k}/2}\right),$
with $Z_k$ denoting the Pauli-Z operator on the $k$th qubit. Note that if there is no data-dependent unitary on qubit $k$ in the $j$'th data-dependent layer, then we simply set $\phi_{jk}(\vx)= 0$ for all $\vx$. We will consider circuits in the form of Eq.~\eqref{eq:standard_expression} to be in \textit{standard form}.

\subsection{Quantum kernels are entangled tensor kernels\label{Sec:quatnumkernelsinTN}}

We now show how, using the standard form for data-dependent quantum circuits discussed in the previous section, we can write quantum kernels in a tensor network form, which allows us to easily identify them as entangled tensor kernels. To this end, we use tensor network manipulations to rewrite evaluation of the kernel as follows:
\begin{widetext}

\begin{align}
K_U(\vx,\vx') &= \abs{\bra{0}^{\otimes n}U^{\dag}(\vx)U(\vx')\ket{0}^{\otimes n}}^2\\
&= \vcenter{\hbox{\includegraphics[height=0.2\linewidth]{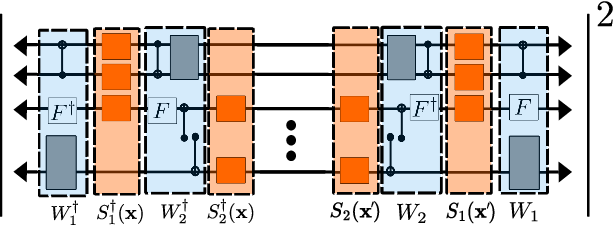}}} \\
&= \vcenter{\hbox{\includegraphics[width=0.5\linewidth]{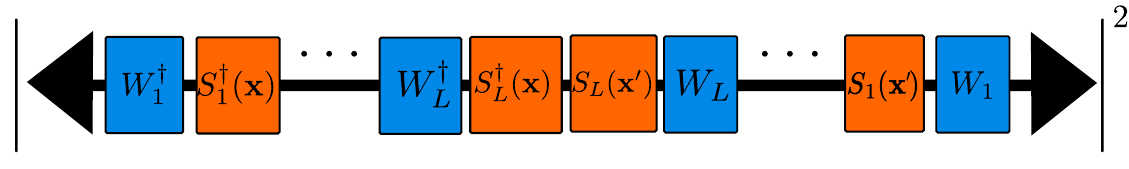}}}\label{eq:thick_line}\\
&= \vcenter{\hbox{\includegraphics[height=0.2\linewidth]{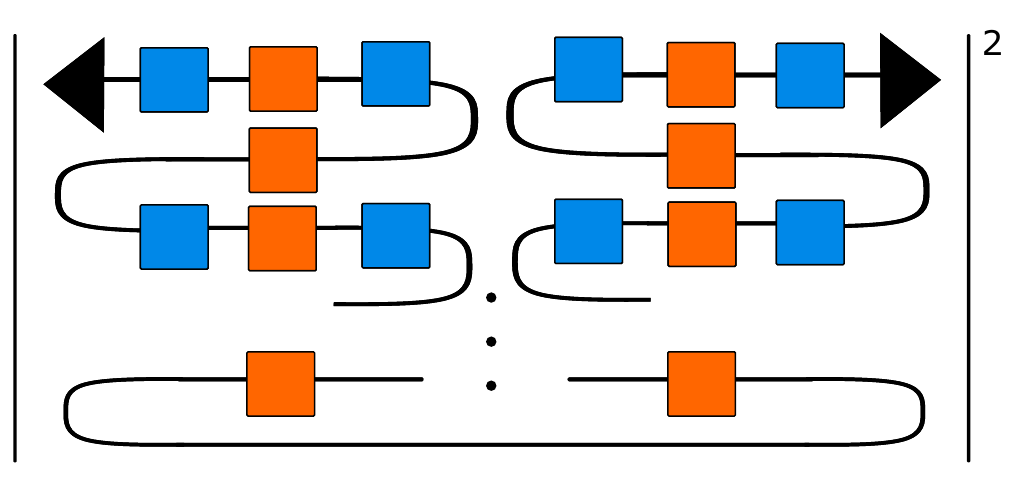}}}\\
&= \vcenter{\hbox{\includegraphics[height=0.2\linewidth]{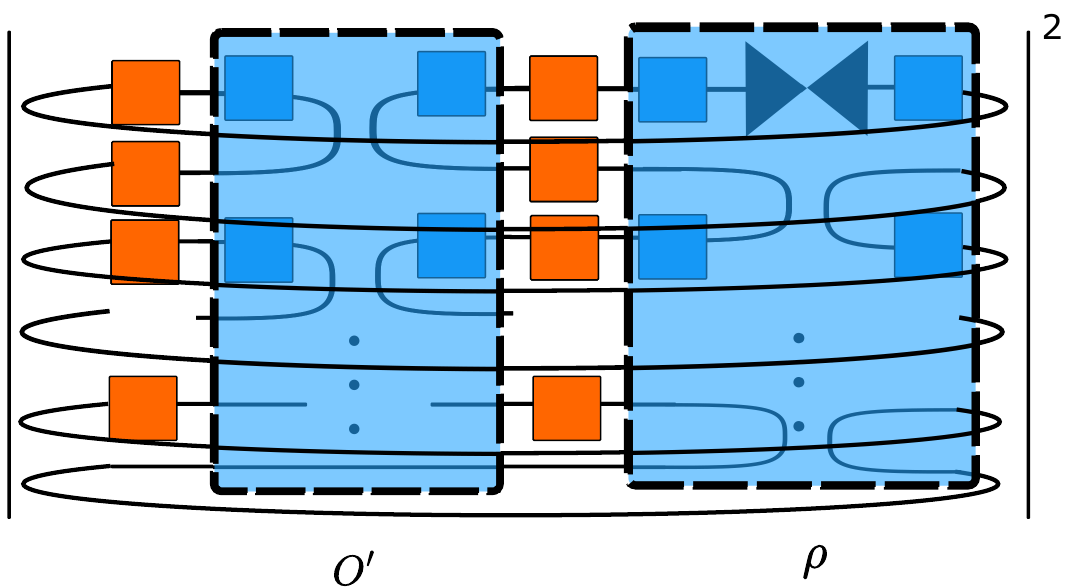}}}\\
&=\vcenter{\hbox{\includegraphics[height=0.11\linewidth]{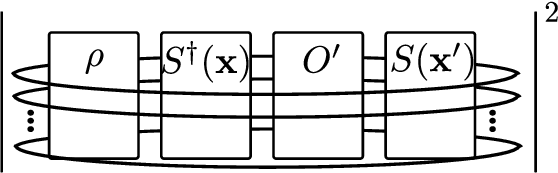}}}\label{eq:trace_diagram}\\
&=\abs{\Tr{\rho S^{\dagger}(\vx)O'S(\vx')}}^2,
\end{align}
where from Eq.~\eqref{eq:thick_line} onwards each line in the diagram represents an $n$-qubit wire, and we have defined
\begin{align}
&O'=\begin{cases}\bigotimes_{j=1}^{(L-1)/2}(W_{2j}^{\dag} \otimes I)\ketbra{\Phi}(W_{2j} \otimes I)\otimes I, \hspace{2em}\text{if L is odd},\\
     \bigotimes_{j=1}^{L/2}(W_{2j}^{\dag} \otimes I)\ketbra{\Phi}(W_{2j} \otimes I)\otimes I, \hspace{3.5em}\text{if L is even},
 \end{cases} \label{eq:generalO}\\
&\rho =\begin{cases}W_1\ketbra{0}W_1^{\dag} \otimes \bigotimes_{j=1}^{(L-1)/2}( I \otimes W_{2j+1} )\ketbra{\Phi} (I \otimes W_{2j+1}^{\dag} ), \hspace{3.5em}\text{if L is odd},\\
        W_1\ketbra{0}W_1^{\dag} \otimes \bigotimes_{j=1}^{(L/2-1)}( I \otimes W_{2j} )\ketbra{\Phi} (I \otimes W_{2j+1}^{\dag} )\otimes \ketbra{\Phi},  \hspace{0.5em}\text{if L is even},
\end{cases} \label{eq:generalrho}\\
&S(\vx)= \bigotimes_{j=1}^{L}S_j(\vx)\\
&\ketbra{\Phi} = \sum_{i,j=1}^{2^n}\ket{ii}\bra{jj}.
\end{align}
\end{widetext}
We then proceed by using the equality
\begin{equation}
    \Tr{D^{\dag}ADB} = \bra{D}\left(A\odot B^\top\right)\ket{D},
\end{equation} where $\odot$ represents the Hadamard (element-wise) product, $D$ is a diagonal matrix, and $\ket{D}$ denotes the vector whose elements are entries of diagonal matrix $D$. Using this equality and the fact that $S(\vx)$ is a tensor product of diagonal matrices, the quantum kernel becomes

\begin{equation}\label{eq:next_step}
\begin{split}
    K_U(\vx,\vx') &= \vcenter{\hbox{\includegraphics[height=0.2\linewidth]{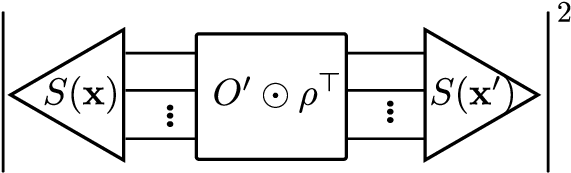}}}\\
&=\vcenter{\hbox{\includegraphics[height=0.2\linewidth]{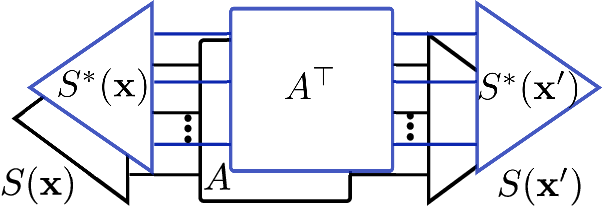}}}\\
&=\vcenter{\hbox{\includegraphics[height=0.2\linewidth]{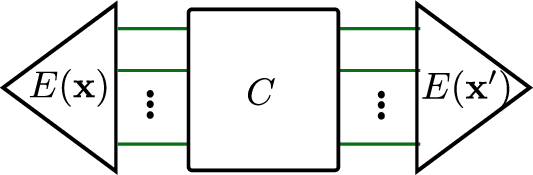}}}.\\
    \end{split}
\end{equation}
In the above we have defined 
\begin{align}
&A := \left(O' \odot \rho^\top\right)\\
&C := A^\top \otimes_v A \\
&E(\vx) := S^{*}(\vx)\otimes_v S(\vx),
\end{align}
where $\otimes_v$ is a ``vertical nearest-neighbor'' tensor product, which juxtaposes two equal dimension tensors ``vertically'' such that each index of the first tensor becomes the nearest neighbor of the corresponding index of the second tensor. Said another way, as illustrated above, one performs $\otimes_v$ by writing one tensor on top, and slightly shifted vertically of the other tensor, so that their legs are interlaced. To make things explicit, for $C$ we have that
\begin{equation}
\begin{split}\label{eq:vertical tensor}
&C_{i_1i_2\ldots i_{2nL};j_1j_2\ldots j_{2nL}} \\&= A^{\top}_{i_1i_3\ldots i_{2nL-1};j_1j_3\ldots j_{2nL-1}}\times A_{i_2i_4\ldots i_{2nL};j_2j_4\ldots j_{2nL}},
\end{split}  
\end{equation}

where one should keep in mind that each black and blue wire in the diagram above in fact represents $n$ 2-dimensional wires. 
Additionally, recalling the definition of $S(\vx)$, and the ordering implied by $\otimes_v$, we see that $E(\vx)$ is now an $nL$ product of $4$-dimensional vectors which is given by,
\begin{equation}
\begin{split}
   \vcenter{\hbox{\includegraphics[height=0.2\linewidth]{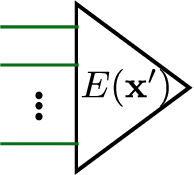}}}&= \bigotimes^L_{j=1}\left(\bigotimes_{k=1}^n \begin{pmatrix}
         1 \\ e^{i\phi_{jk}(\vx')} \\ e^{-i\phi_{jk}(\vx')} \\ 1
     \end{pmatrix}\right) \\&= \bigotimes^L_{j=1}\left(\bigotimes_{k=1}^n \ket{E^{(jk)}(\vx')}\right). 
\end{split}
\end{equation}

Moreover, we observe that the core matrix $C$ is complex-valued, and that complex vector $\ket{E(\vx)}$ possesses many redundant 1's in known indices. We can remove this redundancy by introducing an isometry
\begin{equation}
    P := \frac{1}{\sqrt{2}}\begin{pmatrix}
         1 &0 &0 \\
         0 &1 &i\\
         0 &1 &-i\\
         1 &0 &0
     \end{pmatrix} = \frac{1}{\sqrt{2}}\left(|I\rangle\rangle\bra{0} + |X\rangle\rangle\bra{1} + |Y\rangle\rangle\bra{2}\right),
\end{equation} where for some matrix $M$, we have defined $|M\rangle\rangle$ as its column-major vectorization. From now on, for notational simplicity, we will set $N := nl$ and combine `layer,qubit' index $jk$ into one index $k \in \{1,\ldots,N=nl\}$. This gives us the relation
\begin{align}\label{eq:incorporate_isometry}
    \ket{E^{(k)}(\vx)}&=\begin{pmatrix}
         1 \\ e^{i\phi_{k}(\vx)} \\ e^{-i\phi_{k}(\vx)} \\ 1
     \end{pmatrix} \\&= \sqrt{2}\begin{pmatrix}
         1 &0 &0 \\
         0 &1 &i\\
         0 &1 &-i\\
         1 &0 &0
     \end{pmatrix}\begin{pmatrix}
         \frac{1}{\sqrt{2}} \\ \frac{1}{\sqrt{2}}\cos{\phi_{k}(\vx)}\\ \frac{1}{\sqrt{2}}\sin{\phi_{k}(\vx)}
     \end{pmatrix} \\&:= 2P\ket{T^{(k)}(\vx)}.
\end{align} We have set $\bra{T^{(k)}(\vx)}\ket{T^{(k)}(\vx)} = 1$, for normalization. Additionally, when tensors lose their superscripts, one should understand them as the $N$-fold tensor product of the tensors with the same symbol -- for example, $\ket{T(\vx)} = \bigotimes_{k=1}^N\ket{T^{(k)}(\vx)}$.  Now by exploiting Eq.~\eqref{eq:incorporate_isometry}, we can obtain
\begin{align}
    K_U(\vx,\vx') &= \bra{E(\vx)} C \ket{E(\vx')}\label{eq:EC}\\&=4^N\times\vcenter{\hbox{\includegraphics[height=0.15\linewidth]{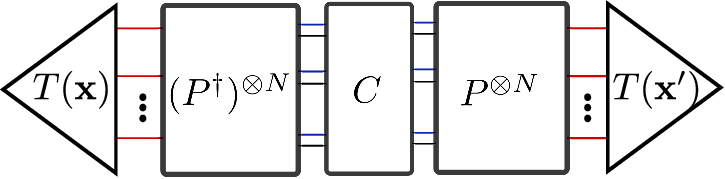}}}\\
    & := \vcenter{\hbox{\includegraphics[height=0.2\linewidth]{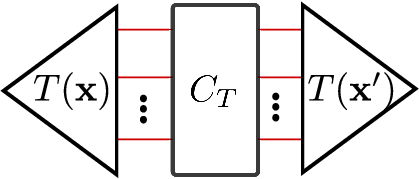}}}\\
    &= \vcenter{\hbox{\includegraphics[height=0.2\linewidth]{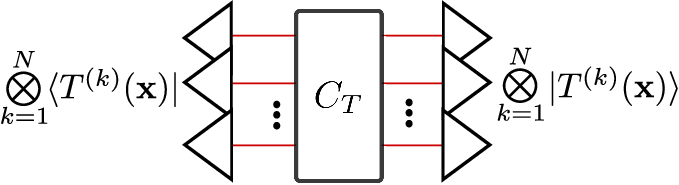}}},\label{eq:canonical form}
\end{align}where $C_T := 4^{N}\left(\bigotimes_k P^{\dag}\right) C\left(\bigotimes_k P\right) $ and now red lines indicate 3-dimensional wires. We added the last equality to clearly show the tensor product structure of $|T(\vx)\rangle$. Given this, we see by comparing Eq.~\eqref{eq:canonical form} with Eq.~\eqref{eq:ETK_TN_form}, that  $K_U$ is indeed an entangled tensor kernel, with local feature maps
\begin{equation}\label{eq:quantumfeaturemap}
    \ket{T^{(k)}(\mathbf{x})} = \frac{1}{\sqrt{2}}\begin{pmatrix}
    1 \\ \cos{\phi_{k}(\mathbf{x})} \\ \sin{\phi_{k}(\mathbf{x})}
    \end{pmatrix},
\end{equation}
and a quantum-circuit generated symmetric PSD core tensor $C_T$. 

Finally, we note that we can also identify
\begin{equation}
    \left(C_T\right)_{\mathbf{i}\mathbf{j}} = 4^{N}\times \frac{1}{2^{N}}\Tr{\mathcal{P}_{\mathbf{i}}A\mathcal{P}_{\mathbf{j}}A}\quad\quad \mathbf{i},\mathbf{j} \in \{0,1,2\}^{N},
\end{equation} where $\mathcal{P}_{\mathbf{i}}$ is the Pauli string associated to index-string $\mathbf{i}$. In light of this, one could say that $C_T$ is a (truncated) re-scaled Pauli transfer matrix (PTM) of the linear map $\mathcal{A} : \mathbf{M} \mapsto A\mathbf{M}A$. Additionally, we note that the matrix $C_T$ is also symmetric and now has real-valued elements.

\section{The landscape of Entangled Tensor Kernels}\label{sec:landscape}

\begin{figure*}[t]
    \includegraphics[width = 1.7\columnwidth]{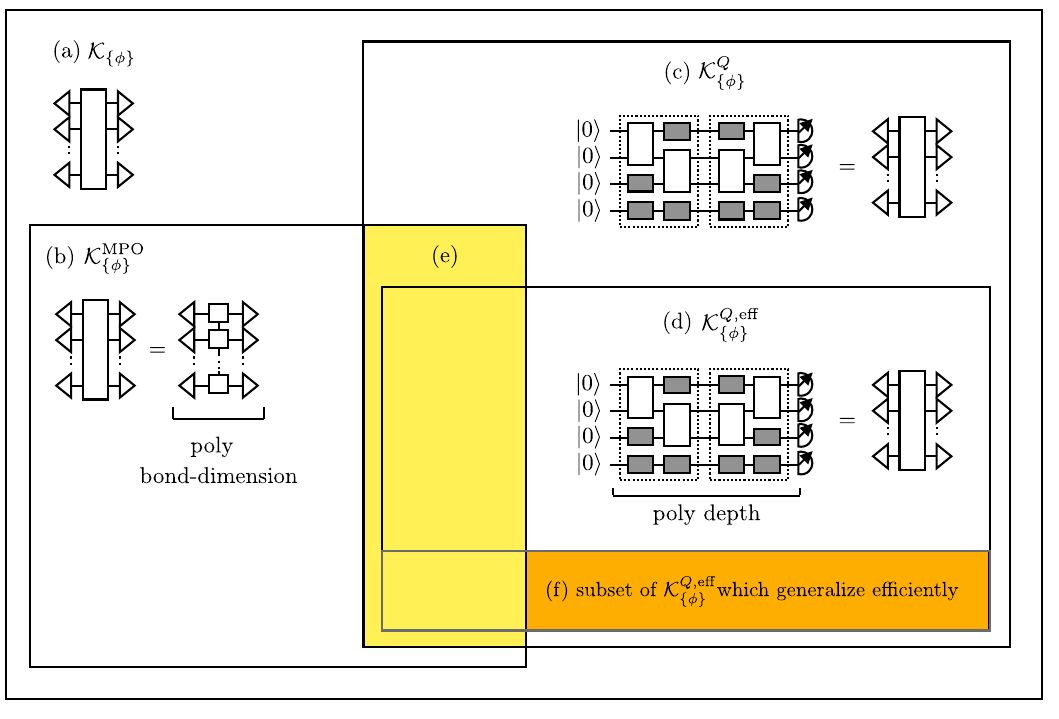}
    \caption{
    An overview of the conjectured landscape of ETKs, for some fixed set of local feature maps derived from data pre-processing functions $\{\phi\}$. Region (a) is the set of \textit{all} such ETKs, and region (b) is the subset of those ETKs whose core tensor admits a decomposition into a polynomial bond-dimension MPO, and can therefore be evaluated efficiently classically (when given this MPO). Region~(c) contains all those ETKs which can be evaluated using an embedding quantum kernel, and region (d) is the subset of those ETKs which can be evaluated using a polynomial depth quantum circuit -- i.e. those which can be evaluated efficiently quantumly. In Section~\ref{sss:inductive_bias} we discuss the \textit{inductive bias} which distinguishes quantum kernels that are in region (d), but not in region (b). Region (f) is the subset of efficient quantum kernels which also generalize efficiently (for some problem of interest). We identify these quantum kernels as naturally interesting quantum kernels for further exploration, and discuss candidates in Section~\ref{ss:summary}. Region (e) is the subset of quantum kernels that can in principle be evaluated efficiently classically. The existence of this set suggests that ETKs with poly-bond-dimension MPOs can sometimes be used to \textit{dequantize} quantum kernels, which is discussed in Section~\ref{ss:classical_surrogates}. Whether or not region (b) is a proper subset of region (c) is an open question. }
    \label{fig:Overview}
\end{figure*}

In light of the results of the previous section, we proceed here to further examine the landscape of ETKs -- illustrated in Figure~\ref{fig:Overview} -- with the goal of better understanding the relationship between quantum kernels and classical ETKs. In particular, we will also highlight in Sections~\ref{ss:highentangledETKs} -~\ref{ss:classical_surrogates} the implications this relationship has for the design of both potentially interesting quantum kernels and ``classical surrogate kernels'' for dequantizing quantum kernel methods.

To this end, we start by noting that the local feature maps $\{|T^{(k)}(\vx)\rangle\}$ of the ETK representation of a quantum kernel are uniquely determined by a set of data pre-processing functions $\{\phi\}:=\{\phi_k\,|\,k \in [N]\}$. With this in mind, we assume some fixed set of pre-processing functions $\{\phi\}$, and start by defining $\mathcal{K}_{\{\phi\}}$, the set of \textit{all} ETKs using the associated local feature maps $\{|T^{(k)}(\vx)\rangle\}$, via
\begin{equation}
\begin{split}
    \mathcal{K}_{\{\phi\}} := \{&K_C\,:\,K_C(\vx,\vx') = \langle T(\vx)|C|T(\vx\rangle\},
\end{split}
\end{equation}where $C$ is positive semidefinite.
As we have discussed in Section~\ref{ss:complexity}, not all ETKs in $\mathcal{K}_{\{\phi\}}$ can be evaluated efficiently. However, we have seen that whenever the core tensor $C$ can be represented by an MPO with polynomial bond-dimension, then the associated ETK \textit{can} be efficiently evaluated classically. In light of this, we define the subset of $\mathcal{K}_{\{\phi\}}$ whose core tensor $C$ admits a decomposition into an MPO with polynomial bond-dimensions as $\mathcal{K}^{\mathrm{MPO}}_{\{\phi\}}$. By construction, all ETKs in $\mathcal{K}^{\mathrm{MPO}}_{\{\phi\}}$ can be evaluated efficiently classically, when given the MPO representation of the core tensor $C$.

We would now like to understand how quantum kernels fit into the picture. To this end, let us denote by $C_T(\{W\},\{\phi\})$ the core tensor $C_T$ which one obtains, via the derivation in Section~\ref{Sec:quatnumkernelsinTN}, when starting from a quantum kernel $K_U$ with some data-dependent quantum circuit $U(\vx) = \prod_{j=1}^L S_j(\vx)W_j$, which uses the pre-processing functions $\{\phi\}$. This allows us to define the set of ``quantum ETKs'' $\mathcal{K}^Q_{\{\phi\}}$ via
\begin{equation}
\begin{split}
    \mathcal{K}^Q_{\{\phi\}} := \left\{K_C\,:\,K_C(\vx,\vx') = \langle T(\vx)|C_T(\{W\},\{\phi\})|T(\vx')\rangle\right\}.   
\end{split}
\end{equation}
In words, $\mathcal{K}^Q_{\{\phi\}}$ is nothing but the set of entangled tensor kernels one can realize by using quantum kernels $K_U$ with valid data-dependent quantum circuits $U(\vx) = \prod_{j=1}^L S_j(\vx)W_j$, using pre-processing functions $\{\phi\}$. As any ``quantum generated'' core tensor $C_T(\{W\},\{\phi\})$ is PSD, we immediately have that
\begin{equation}
K^{Q}_{\{\phi\}} \subseteq K_{\{\phi\}},
\end{equation}
i.e. as illustrated in Figure~\ref{fig:Overview}, ``quantum ETKs'' are a subset of all possible ETKs using the same local feature maps. 

Consider now some ETK $K_C\in\mathcal{K}^Q_{\{\phi\}}$. By construction, we know that there exists a data-dependent unitary $U(\vx)$, such that $K_C=K_U$. As such, one could evaluate $K_C$ by executing the quantum circuit shown in Figure~\ref{fig:QK_evaluation}. When implenting $K$ via this route, the efficiency of the implementation of the kernel depends on the depth of the quantum circuit $U(\vx)$ -- when this circuit can be efficiently implemented, then the kernel can be evaluated efficiently. Motivated by this observation, we call the subset of quantum ETKs $\mathcal{K}^Q_{\{\phi\}}$ which can be efficiently evaluated on a quantum computer $\mathcal{K}^{Q,\mathrm{eff}}_{\{\phi\}}$. 

To summarize, at this stage we have established the following relationships, also illustrated in Figure~\ref{fig:Overview}:
\begin{align}
\mathcal{K}^{Q,\mathrm{eff}}_{\{\phi\}} \subset \mathcal{K}^{Q}_{\{\phi\}} \subseteq &\,\mathcal{K}_{\{\phi\}} \nonumber\\
&\,\,\rsubset\\ 
&\mathcal{K}^{\mathrm{MPO}}_{\{\phi\}}.\nonumber
\end{align}
With this in hand, we would now ideally like to understand the relationships between the set of quantum kernels which can be evaluated efficiently with a quantum computer $\mathcal{K}^{Q,\mathrm{eff}}_{\{\phi\}}$, and the set of ETKs which can be evaluated efficiently classically via tensor network contractions $\mathcal{K}^{\mathrm{MPO}}_{\{\phi\}}$. We conjecture that the relationships between these sets are as illustrated in Figure~\ref{fig:Overview}. With this in mind, the following questions are of particular interest to us:

\begin{enumerate}
    \item Do there exist kernels in $\mathcal{K}^{Q,\mathrm{eff}}_{\{\phi\}}$ that are \textit{not} in $\mathcal{K}^{\mathrm{MPO}}_{\{\phi\}}$? Said another way: Are there quantum kernels which can be evaluated efficiently with a quantum computer, but do not admit efficient evaluation via the ETK perspective? If the answer to this question is ``Yes'', then what properties do these kernels have? Are they potentially useful? We explore this question and provide some (partial) answers in Sections~\ref{ss:highentangledETKs} and~\ref{ss:summary}.
    \item Similarly, are there kernels in $\mathcal{K}^{\mathrm{MPO}}_{\{\phi\}}$ that are \textit{not} in $\mathcal{K}^{\mathrm{Q,eff}}_{\{\phi\}}$? We conjecture that the answer to this question is ``Yes'', but leave a resolution of this to future work. 
    \item Finally, given the existence of kernels in the intersection $\mathcal{K}^{Q,\mathrm{eff}}_{\{\phi\}}\cap\mathcal{K}^{\mathrm{MPO}}_{\{\phi\}}$, to what extent can we use this to \textit{dequantize} quantum kernel methods. We discuss the potential and limitations in this regard in Section~\ref{ss:classical_surrogates}.
\end{enumerate}

 Before proceeding, we comment on projected quantum kernels. Projected quantum kernels introduced in Ref.~\cite{huang2021power} use classical functions of quantities derived from reduced states, so they do not directly fit our ETK construction in full generality. However, when a projected kernel can be written as an embedding kernel on an enlarged feature map (e.g., by explicitly including the measured local features), the ETK lens may still apply to that embedding representation. Moreover, recent works have provided unifying perspectives~\cite{gan2023unified,Gil_Fuster_2024}; in particular, it has been shown that \textit{any} reasonable quantum kernel can be written as an embedding quantum kernel~\cite{Gil_Fuster_2024}. We leave a systematic treatment of projected kernels in the ETK language to future work.

\subsection{Quantum kernels can potentially realize super-polynomial bond-dimension ETKs}\label{ss:highentangledETKs}

As illustrated in Figure~\ref{fig:Overview}, there is the possibility of quantum kernels which:
\begin{enumerate}
\item Can be evaluated efficiently via the execution of a polynomial depth quantum circuit.
\item Do not admit a polynomial bond-dimension MPO decomposition for the core tensor of the ETK representation.
\end{enumerate}
We refer to such quantum kernels as having ``super-polynomial bond-dimension''. As per the discussion in Section~\ref{ss:complexity}, such kernels cannot be evaluated efficiently classically by simply contracting the tensor network defining the ETK representation. As such one can think of ``super-polynomial bond-dimension'' as an \textit{inductive bias} which can be possessed by efficient quantum kernels, but not by ETKs which are classically efficient to evaluate via the tensor network contraction method discussed in Section~\ref{ss:complexity}.

We stress however that having super-polynomial bond-dimension is a \textit{necessary but not sufficient} condition for a quantum kernel to be hard to evaluate classically. We cannot rule out the existence of ETKs with super-polynomial bond-dimension, which can nevertheless be evaluated efficiently classically via some other method than contracting the tensor network representation of the ETK. In this sense, we can view ``super-polynomial bond-dimension'' as  an inductive bias which is \textit{easily} accessible to efficient quantum kernels, but not so obviously accessible to kernels which are efficient to evaluate classically. Given this, we leave to future work the following questions:

\begin{enumerate}
\item How can we design quantum kernels to ensure that they have super-polynomial bond-dimension?
\item Can one construct and analyze examples of super-polynomial bond-dimension ETKs, which can nevertheless be evaluated efficiently classically?
\end{enumerate}

\subsection{Do useful quantum kernels exist?}\label{ss:summary}

In the previous section, we discussed how having super-polynomial bond-dimension is a necessary but not sufficient condition for a quantum kernel to be hard to evaluate classically. However, for a quantum kernel to be potentially \textit{useful} for a specific problem, one should both (a) provide evidence that the quantum kernel cannot be evaluated efficiently classically, and (b) provide evidence that the kernel generalizes efficiently for some interesting problem -- i.e. that it can generalize well on a polynomially sized dataset of interest. As having super-polynomial bond-dimension provides some evidence for being hard to evaluate efficiently classically, we ask here whether there are quantum kernels that have a super-polynomial bond-dimension \textit{and} generalize efficiently.

As discussed in Refs.~\cite{Kubler2021inductive,Canatar2021spectral,canatar2023bandwidth}, one way to show that a kernel generalizes efficiently (for some set of problems) is to show that the spectrum of the kernel is concentrated on a polynomially sized subset of eigenvalues. At a high-level, such kernels will generalize efficiently for target functions that are supported on the corresponding eigenfunctions. At first glance, one might think that this immediately precludes the existence of quantum kernels which both have super-polynomial bond-dimension and generalize efficiently! Indeed, one might think that when the spectrum of the kernel is concentrated on polynomially many eigenvalues, one could simply truncate the size of the feature space to some polynomial in the dimension of the data (namely, the feature space spanned by the top eigenfunctions) and therefore evaluate the kernel efficiently classically (by exploiting the truncated core tensor that would admit polynomial bond-dimension).

However, here we show that this is \textit{not} the case, and that there are indeed kernels which both have super-polynomial bond-dimension and generalize efficiently. We stop short of the true goal of showing the existence of \textit{quantum} kernels with these properties, but discuss why one might expect them to exist. 

To begin, consider the kernel
\begin{equation}
\begin{split}
    K_G(\vx,\vx') = \bra{T(\vx)} U^{\dag} D U \ket{T(\vx')},
\end{split}
\end{equation} 
where $D$ has only $O(\text{poly}(d))$ nonzero diagonal elements, or its values are concentrated on $O(\text{poly}(d))$ elements, where $d$ is the dimension of the local feature maps. The unitary $U$, however, has $O(\text{exp}(d))$ bond-dimensions when represented as an MPO. Using the tools and techniques of Refs.~\cite{Kubler2021inductive,Canatar2021spectral,canatar2023bandwidth}), one can see that $K_G$ will indeed generalize efficiently (for some set of problems). However, the contracted core tensor $C_G := U^{\dag}DU$ could have $O(\text{exp}(d))$ bond-dimension, placing it outside $\mathcal{K}^{\mathrm{MPO}}_{\{\phi\}}$. As such, this kernel has both a super-polynomial bond-dimension \textit{and} generalizes efficiently for some class of problems. Whether $\mathcal{K}^{\mathrm{Q,eff}}_{\{\phi\}}$ -- the set of \textit{quantum} kernels which can be evaluated efficiently -- indeed includes kernels like $K_G$ remains open. However we conjecture it does, due to the ability of such kernels to generate super-polynomial bond-dimension ETKs.

We note that one can interpret $K_G$ as a model utilizing a (relatively) small function space spanned by eigenfunctions that are highly entangled (in the sense discussed in Section~\ref{sec:ETKs}), so that they are classically hard to compute (at least via tensor network contractions). These classically hard eigenfunctions appear as exponentially large sums of trigonometric functions,  $\hat{e}_i(\vx) = \sum_jU_{ij}T_j(\vx)$. Even if the form of these eigenfunctions is explicitly known (i.e. one knows all $U_{ij}$s), evaluating them is nontrivial because the summation involves contracting high-bond-dimensional tensor networks, which lack concise representations. Thus, we can characterize candidates for \textit{usefu}l quantum kernels as those with the following properties:
\begin{enumerate}
    \item Generalize efficiently: The spectrum of the kernel is concentrated on polynomially many eigenvalues, and therefore the kernel method is biased towards functions spanned by a polynomial number of top eigenfunctions. 
    \item Classically hard eigenfunctions: The top eigenfunctions that span the (polynomially large) effective function space cannot be efficiently evaluated using only the given description of the data-dependent circuit $U(\vx)$.
\end{enumerate} 
One final important remark is that the above conditions are only \textit{necessary} (but not sufficient) for useful quantum kernels. More specifically, we cannot rule out the possibility that a clever classical algorithm (different from just contracting the given tensors) might exist to evaluate super-polynomial bond-dimension ETKs efficiently.

\subsection{ETKs as classical surrogates for quantum kernels}\label{ss:classical_surrogates}

A standard approach to assessing whether or not one can gain a quantum advantage from a specific quantum algorithm, is to ask whether one can use insights into the nature and structure of the quantum algorithm to construct a purely classical algorithm which is both efficient, and guaranteed to perform as well as the quantum algorithm. In other words, to ask whether one can \textit{dequantize} the quantum algorithm~\cite{tang2022dequantizing}. For those quantum machine learning algorithms, including quantum kernel methods, which work by optimizing over a model class defined by a parameterized quantum circuit, a natural approach to dequantization is to look for a classical model class whose \textit{inductive bias} is similar to that of the quantum model class, and can be evaluated efficiently classically. Such a model class has come to be known as a \textit{classical surrogate} for the quantum model class, and surrogates for variational QML models for supervised learning have recently begun to be proposed and studied~\cite{Schreiber_2023,landman2022classicallyapproximatingvariationalquantum, Sweke2023potential}. 

From this perspective, the fact that all quantum kernels are ETKs immediately suggest ETKs as a natural classical surrogate for quantum kernels. More specifically, as we have seen in Figure~\ref{fig:Overview}, there exist quantum kernels $K_U$, which are also in $K^{\mathrm{MPO}}_{\{\phi\}}$, and can therefore be evaluated efficiently classically, when given the MPO description of the ETK core tensor. This seems to immediately suggest that kernel methods using such quantum kernels can be immediately dequantized, and cannot be used to obtain any quantum advantage -- even if the circuit $U(x)$ is classically hard to simulate! Specifically, instead of using the quantum circuit to evaluate the kernel, one could just efficiently evaluate the corresponding ETK classically via tensor network contractions. Unfortunately however, there is the following subtlety:
\begin{center}
    \textit{Given a description of a data-dependent quantum circuit $U$, we cannot always efficiently obtain the optimal MPO description of the core tensor $C_T(\{W\},\{\phi\})$. }
\end{center}
As such, even if we are promised that a quantum kernel has a polynomial bond-dimension MPO core tensor, we might not be able to efficiently obtain this MPO from the circuit description that we are given. 

In light of this, there are three natural strategies to dequantization of quantum kernel methods via ETKs as classical surrogates:

\begin{enumerate}
\item Assess whether the structure of the quantum circuit $U$ allows for efficient extraction of an MPO representation of the corresponding ETK core tensor. 
\item \textit{Learn} a polynomial bond-dimension positive semidefinite MPO core tensor (via optimization of kernel-target alignment for example~\cite{Hubregtsen_2022}), and use the corresponding ETK.
\item Use the structure of the quantum circuit $U$ to make informed guess, or random selection, of a polynomial bond-dimension positive semidefinite MPO core tensor, and use the corresponding ETK (in some ways, analogously to the use of Random Fourier Features for dequantizing variational QML~\cite{landman2022classicallyapproximatingvariationalquantum,Sweke2023potential}).
\end{enumerate}
We leave further study of these approaches to dequantization of classical kernel methods to future work, however we make a few remarks. Firstly, as per Section~\ref{ss:ETK}, in order to be a valid kernel, we know that the core tensor of an ETK needs to be positive semidefinite. As such, as we have discussed in Section~\ref{ss:complexity}, a natural approach when trying to guess or learn a suitable ETK core MPO is to use \textit{locally-purified} MPOs (LP-MPOs), which are positive semidefinite by construction. Additionally, one might think that only those quantum kernels which are also in $\mathcal{K}^{\mathrm{MPO}}_{\{\phi\}}$ are susceptible to dequantization via the approaches described above. However, this is not necessarily the case. Indeed, even if the optimal MPO representation of the ETK core tensor for a quantum kernel requires super-polynomial bond-dimension, it may be the case that a polynomial bond-dimension MPO provides a suitably accurate approximation. Moreover, and counter-intuitively, the fact that ETKs in $\mathcal{K}^{\mathrm{MPO}}_{\{\phi\}}$ are more constrained in certain ways than quantum kernels, may indeed give them advantages in generalization over many quantum kernels (as discussed in Ref.~\cite{Sweke2023potential}).

\section{The power of the ETK lens for analysis of quantum kernels}\label{ss:example_onelayer}

As we have discussed briefly in Section~\ref{sss:inductive_bias}, obtaining the Mercer decomposition of a kernel allows one to extract a variety of quantitative insights into the inductive bias, sample complexity and generalization capacity of the kernel. Indeed, Refs.~\cite{Kubler2021inductive,Shaydulin_2022, canatar2023bandwidth} have all obtained a variety of quantitative insights into properties of quantum kernels, via Mercer decompositions of simple examples. These insights have then facilitated concrete recommendations for both the design of quantum kernels, and suitable applications.

\subsection{Analysis on single-layer model with arbitrary n-qubit unitary}

In this section, in order to showcase the power of the ETK lens for the analysis of quantum kernels, we show how to use the ETK picture to obtain the Mercer decomposition of a family of specific quantum kernels, which goes beyond those previously studied in Refs.~\cite{Kubler2021inductive,Shaydulin_2022,canatar2023bandwidth}. In particular, previous works have analyzed what we refer to in this work as \textit{product kernels}. In this section, we will show how to use the insights from Section~\ref{Sec:quantumkernelsareETK}, together with techniques from Appendix~\ref{app:mercer_decomp}, to obtain the Mercer decomposition of a specific family of quantum kernels.

More specifically, we study quantum kernels $K_U$ on $n$ qubits, satisfying the following two assumptions:

\textbf{Assumption 1: Single-layer of data-dependent gates.} We assume that the data-dependent unitary specifying the quantum kernel is of the form
\begin{align}
U(\vx) &= S(\vx)W\\
&= \left(\bigotimes_{k=1}^{n}e^{-i\phi_{k}(\vx)Z_{k}/2}\right)W.
\end{align}
This is a special case of the general setting studied in Section~\ref{Sec:quantumkernelsareETK}, with $L=1$.

\textbf{Assumption 2: Simple pre-processing.} We assume that $\vx\in [-\pi,\pi]^n$, and that ${\phi_k(\vx) = \vx_k}$. 

With these assumptions, we start by deriving the ETK representation of the kernel, as per the techniques of Section~\ref{Sec:quantumkernelsareETK}. To this end, we see from Eqs~\eqref{eq:generalO} and~\eqref{eq:generalrho}, that when $L=1$, one has  $O' = I$, and $\rho = W\ketbra{0}^{\otimes n}W^\dag := \ketbra{\psi}$. We then have from Eq.~\eqref{eq:next_step} that
\begin{equation}
    K_{U}(\vx,\vx') = \bra{\diag(S(\vx))} \left(I \odot \ketbra{\psi}\right)  \ket{\diag(S(\vx'))} \times(c.c).
\end{equation} 
In this case, when we apply the Hadamard product with the identity matrix $I$, it selects the diagonal elements, leading to  $(I \odot \ketbra{\psi}) = \diag(\ketbra{\psi}) := D(\psi^2)$, which is a diagonal matrix whose elements are the squared magnitudes of the components of $\ket{\psi}$. Using this in Eq.~\eqref{eq:EC}, we obtain
\begin{align}
     &K_{U}(\vx,\vx')= \bra{E(\vx)}C\ket{E(\vx')} 
     \\&= \left(\begin{pmatrix}
         1 \\ e^{-i\phi_k(\vx)} \\ e^{i\phi_k(\vx)} \\ 1
     \end{pmatrix}^{\otimes n}\right)^\top (D(\psi^2) \otimes_v D(\psi^2))\begin{pmatrix}
         1 \\ e^{i\phi_k(\vx')} \\ e^{-i\phi_k(\vx')} \\ 1
     \end{pmatrix}^{\otimes n} \label{eq:onelayercomplex}
\end{align}

We have now arrived at an ETK representation of the quantum kernel. In Section~\ref{Sec:quantumkernelsareETK} we proceeded to use an isometry $P$ to remove redundant components of $|E(\vx)\rangle$ and obtain a simplified ETK. However, as obtaining the Mercer decomposition requires the orthonormalization of components $|E(\vx\rangle)$, and the removal of redundant components, we can simply proceed with the Mercer decomposition from here.  To this end, we recall from Appendix~\ref{app:mercer_decomp} that the first step is to obtain an orthonormal basis for the component functions of $|E(\vx)\rangle$. However, we note that under Assumption 2 above, we have that 
\begin{equation}
|E(\vx)\rangle = \begin{pmatrix}
         1 \\ e^{-i\vx_k} \\ e^{i\vx_k} \\ 1,
     \end{pmatrix}^{\otimes n}
\end{equation}
and as such all the component functions of $|E(\vx)\rangle$ are already orthonormal with respect to the inner product $\bra{f}\ket{g} = \prod_{k=1}^N\left(\frac{1}{2\pi}\int_{-\pi}^{\pi} f^*(\vx_k)g(\vx_k) d\vx_k\right)$. Indeed, this was the motivation for Assumption 2. In light of this, we can proceed with Step 2 from Appendix~\ref{app:mercer_decomp}. For orthonormalizing tensor $L$, we choose 
\begin{equation}
    L=  \begin{pmatrix}
        1 & 0 & 0 & 0\\
        0 & 1 & 0 & 0\\
        0 & 0 & 1 & 0\\
        -1 & 0 & 0 & 1         
    \end{pmatrix}^{\otimes n},\quad\quad L^{-1}=  \begin{pmatrix}
        1 & 0 & 0 & 0\\
        0 & 1 & 0 & 0\\
        0 & 0 & 1 & 0\\
        1 & 0 & 0 & 1         
    \end{pmatrix}^{\otimes n}. 
\end{equation}which satisfies
\begin{equation}
    L\ket{E(\vx)}=  \begin{pmatrix}
        1 \\
        e^{-i\vx_k} \\
        e^{i\vx_k} \\
        0         
    \end{pmatrix}^{\otimes n}. 
\end{equation}With this, we then have that
\begin{equation}
    \begin{split}
        K_U(\vx,\vx') &= \left(\begin{pmatrix}
        1 \\
        e^{i\vx_k} \\
        e^{-i\vx_k} \\
        0         
    \end{pmatrix}^{\otimes n}\right)^{\top}(L^{-1})^\top CL^{-1}\begin{pmatrix}
        1 \\
        e^{-i\vx'_k} \\
        e^{i\vx'_k} \\
        0         
    \end{pmatrix}^{\otimes n} \\
    &:=\bra{\Bar{E}(\vx)}(L^{-1})^\top CL^{-1}\ket{\Bar{E}(\vx')}.
    \end{split}
\end{equation} Here $C$ is diagonal and elements are
\begin{equation}\label{eq:singlelayerC}
    C_{\mathbf{i};\mathbf{i}} :=C_{i_1i_2,\ldots,i_{2N};i_1i_2,\ldots,i_{2N}} = \psi^2_{i_1i_3,\ldots,i_{2N-1}} \psi^2_{i_2i_4,\ldots,i_{2N}},
\end{equation}with $\mathbf{i} \in \{00,01,10,11\}^{n}$.

Now we will see how conjugating $L$ affects the elements of $C$. Let us denote $\hat{C} := (L^{-1})^\top CL^{-1}$.
\begin{itemize}
    \item For elements of $\hat{C}_{\mathbf{h};\mathbf{h}}$, where index $\mathbf{h}$ contains the sequence $h_{2k-1}h_{2k} = 00$ for any $k\in[N]$, these elements are the sum of elements $C_{\mathbf{i};\mathbf{i}}$ where the index $ \mathbf{i}$ contains $i_{2k-1}i_{2k} = 11$ or $i_{2k-1}i_{2k} = 00$ for the same $k$ such that $h_{2k-1}h_{2k} = 00$
    \item For indices $\mathbf{i}$ not containing sequences of $00$, $ \hat{C}_{\mathbf{i},\mathbf{i}} = C_{\mathbf{i},\mathbf{i}}$.
    \item Non-diagonal elements are created, and the associated indices contain $11$ sequences.
\end{itemize}

The next thing we need to do is a truncation where $\Bar{E}_\mathbf{i}(\vx) = 0$. The zero elements occur whenever there is a $11$ sequence in the index $\mathbf{i}$, so we remove rows and columns where indices have $11$ sequences, and this removes all the non-diagonal elements of $\hat{C}$, giving us
\begin{equation}
\begin{split}
    K_U(\vx,\vx') &= \left(\begin{pmatrix}
        1 \\
        e^{i\vx_k} \\
        e^{-i\vx_k} \\     
    \end{pmatrix}^\top\right)^{\otimes n}\Tilde{C}\begin{pmatrix}
        1 \\
        e^{-i\vx'_k} \\
        e^{i\vx'_k} \\         
    \end{pmatrix}^{\otimes n} \label{eq:mform} \\
    &= \sum_{\mathbf{\alpha} \in \{00,01,10\}^n} \Tilde{C}_{\mathbf{\alpha},\mathbf{\alpha}}e^{-i\omega^{\mathbf{\alpha}}\cdot (\vx - \vx')},
    \end{split}
\end{equation}where frequency vectors $\omega^{\mathbf{\alpha}} \in \{-1,0,1\}^n$ are defined as
\begin{equation}
    (\omega^{\mathbf{\alpha}})_k = \begin{cases}
        0~~\text{if}~~\alpha_{2k-1}\alpha_{2k} = 00\\
        1~~\text{if}~~\alpha_{2k-1}\alpha_{2k} = 01\\
        -1~~\text{if}~~\alpha_{2k-1}\alpha_{2k} = 10\\
    \end{cases}  
\end{equation}
\begin{equation}\label{eq:C_alpha}
    \Tilde{C}_{\mathbf{\alpha},\mathbf{\alpha}}=
       \sum_{\mathbf{\alpha}\in \mathcal{I}_\alpha}\psi^2_{\alpha_1\alpha_3,\ldots,\alpha_{2n-1}} \psi^2_{\alpha_2\alpha_4,\ldots,\alpha_{2n}}, 
\end{equation} and
$\mathcal{I}_{\alpha}$ is the set of all indices containing 00 and 11 sequences where $\alpha_{2k-1}\alpha_{2k}=00$. For example, $\mathcal{I}_{0000} = \{0000,0011,1100,1111\}$, $\mathcal{I}_{0001} = \{0001,1101\}$,  $\mathcal{I}_{0101} = \{0101\}$ and so on. Therefore, the cardinality of the index set becomes $|\mathcal{I}_{\alpha}| = 2^{|\alpha|_{00}}$, where $|\alpha|_{00}$ denotes the number of $00$ sequences in $\alpha$.

We have now finally in Eq.~\eqref{eq:mform} arrived at the Mercer decomposition of $K_U$. Indeed, the diagonal elements of $\tilde{C}$ now contain the eigenvalues of the kernel, from which we can extract insights regarding generalization capacity and sample complexity of learning. 

To this end, it is useful to first make several remarks on the matrix $\Tilde{C}$, and the eigenvalues that it contains. First of all, we note that $\Tilde{C}$ has a `Hermiticity' property. Let $\Bar{\alpha}$ be the index with $01$ and $10$ sequences interchanged. Then we have $\Tilde{C}_{\alpha,\alpha} = \Tilde{C}_{\Bar{\alpha},\Bar{\alpha}}$, and $\omega^{\alpha} = \omega^{\Bar{\alpha}}$. Therefore, $K_U$ is indeed real and can be written in the form of
\begin{equation}
    K_U(\vx,\vx') = \Tilde{C}_{\mathbf{0},\mathbf{0}} + \sum_{\alpha \in \Omega^+}2\Tilde{C}_{\alpha,\alpha}\cos(\omega^{\alpha}\cdot(\vx-\vx')),
\end{equation} where $\Omega^+$ is the set of $\alpha$ indices, obtained by retaining only one index between $\alpha$ and $\Bar{\alpha}$ in $\{00,01,10\}^n$. Secondly, the redundancy of the frequency $\omega^{\alpha}$ is given by $2^{|\alpha|_{00}}$, where $|\alpha|_{00}$ is the number of $00$ sequence in $\alpha$. This results in eigenvalues that tend to concentrate more on indices with many $00$ sequences as they have more terms to be added. In other words, low-degree eigenfunctions (cosine functions associated with the low-Hamming weight frequency $\omega^{\alpha}$) get larger eigenvalues.
With these insights, we can now examine the following two cases:


\textbf{Case A: Approximately uniform pre-encoding state $\mathbf{|\psi\rangle}$.} In this case, we assume that the pre-encoding state $\ket{\psi} = W\ket{\mathbf{0}}$ is approximately uniform -- i.e. all components of $\psi^2$ are on the order of $1/2^n$. We note that this will typically be the case when $W$ is chosen randomly. One quick observation in this case, is that

the largest eigenvalue of the kernel $\tilde{C}_{\mathbf{0},\mathbf{0}}$, associated with the function $\mathbf{1}$, will scale as $2^n(1/4^n) = 1/2^n$. The next largest eigenvalues scale as $2^{n-1}(1/4)^n = 1/2^{n+1}$, and they are associated with the $n$ degree-one frequencies like $\omega = \{(1,0,\ldots,0),\ldots,(0,0,\ldots,1)\}$ and so on. Essentially, we see that the largest eigenvalue decays exponentially. Using the insights from Refs.~\cite{Kubler2021inductive,Canatar2021spectral,canatar2023bandwidth} we can then immediately conclude that such kernels are not able to generalize well, from only a polynomial number of data samples (we refer to the above mentioned references for precise statements).

\textbf{Case B: $(s,\epsilon)$-concentrated pre-encoding state $\ket{\psi_{(s,\epsilon)}}$.} We say a state $\ket{\psi}$ is $(s,\epsilon)$ concentrated, when $\sum_{\alpha\in S}|\psi_\alpha|^2 \geq (1-\epsilon)$ for some index set with $|S| = s$. We denote an $(s,\epsilon)$ concentrated state as $\ket{\psi_{(s,\epsilon)}}$, and use $W_{(s,\epsilon)}$ for unitaries that produce $(s,\epsilon)$ concentrated states.

Physically, sparse---$\epsilon =0$ concentrated---unitaries can be realized by evolution under block-diagonal Hamiltonians, which yield block-diagonal unitaries. Accordingly, one can construct quantum circuits with appropriate symmetries to generate sparse unitaries. Moreover, if the locations of the nonzero matrix elements can be computed efficiently, these unitaries can be implemented efficiently~\cite{Jordan2009efficient}. As an alternative route to concentration, one might use peaked quantum circuits~\cite{aaronson2024verifiable}, which generate states whose maximum amplitude is larger than $1/\text{poly}(n)$.

For this case B, let the amplitude squares of $\ket{\psi_{(s,\epsilon)}}$ in the set $S$ be $L \sim \frac{1-\epsilon}{s}$. Then, according to Eq.~\eqref{eq:C_alpha}, the largest eigenvalue is $\Omega(L^2) = \Omega(1/s^2)$. This is because there exists at least one index set $\mathcal{I}_\alpha$ that contains $|\psi_{\alpha_1\alpha_3,\ldots}^2\psi^2_{\alpha_2\alpha_4,\ldots}| = \Omega(L^2)$. Moreover, note that the number of $\alpha$'s that make $|\psi_{\alpha_1\alpha_3,\ldots}^2\psi^2_{\alpha_2\alpha_4,\ldots}| \sim L^2$ is at most $s^2$, therefore, the number of leading $\tilde{C}_{\alpha,\alpha}$ terms can be at most $s^2$, which means that the eigenspectrum from the $(s,\epsilon)$ concentrated state is at most concentrated as $(s^2, 1-(1-\epsilon)^2)$.

In light of this, we can conclude that in order to be useful -- in the sense of being able to generalize well from polynomially many samples, as discussed in Section~\ref{ss:summary} -- the unitary $W$ which specifies the kernel needs to be such that the pre-encoded state $\ket{\psi} = W\ket{\mathbf{0}}$ has its amplitudes concentrated on only $O(\text{poly}(n))$ basis states. However, the core tensor $C_T$ constructed from such states can (in principle) be approximated by an MPO with $O(\text{poly}(n))$ bond-dimension by truncating exponentially small values $\Tilde{C}_{\alpha,\alpha}$ and retaining only $O(\text{poly}(n))$ significant indices $\alpha$. Furthermore, in this scenario, all eigenfunctions of the kernel can be efficiently computed classically, therefore, any generalizable quantum kernel satisfying assumptions 1 and 2 admit a description from which they can be evaluated efficiently classically!

However, similarly to what we have discussed previously in Section~\ref{ss:classical_surrogates}, this efficient classical evaluation relies on knowing the poly-sized set of dominant eigenvalues. As such, if one can efficiently derive an explicit expression for $\ket{\psi}$ (or more precisely $(\psi^2)$) that allows for the efficient truncation described above, then the kernel can be efficiently dequantized. As we have discussed in Section~\ref{ss:classical_surrogates} though, in general the problem of constructing an efficient representation of $\ket{\psi}$ is hard, but for many circuits for $W$ this may indeed be possible. 

In summary, we see that our analysis in this section has allowed us to clearly link properties of the quantum kernel $K_U$ (satisfying assumptions 1 and 2) to properties of the data-independent unitary matrix $W$ and corresponding pre-encoded state $W\ket{0}$. Eigenvalues of kernel integral operators are determined by the squared amplitudes of pre-encoded states. We have derived an explicit formula to connect them and this is a novel observation that could be made through the lens of ETK. These insights, can hopefully be used to guide the design of quantum kernels.

\subsection{Numerical experiments}\label{ss:numerics}

To complement our analysis and discussions above, here we present numerical experiments comparing single-layer quantum kernels from Haar random unitaries $W_{H}$ (i.e. Case A) and $(s,\epsilon)$-concentrated unitaries $W_{(s,\epsilon)}$ (i.e. Case B). In all simulations we set $\epsilon=0.001$, and to avoid notational clutter, we will omit the explicit dependence on $\epsilon$ when referring to $(s,\epsilon)$-concentrated models. Also, we call $s$-concentrated models $s$-sparse models even though they are only approximately sparse.

\subsubsection{Eigenvalue scaling}

\begin{figure}
    \centering
    \includegraphics[width= 1.0\linewidth]{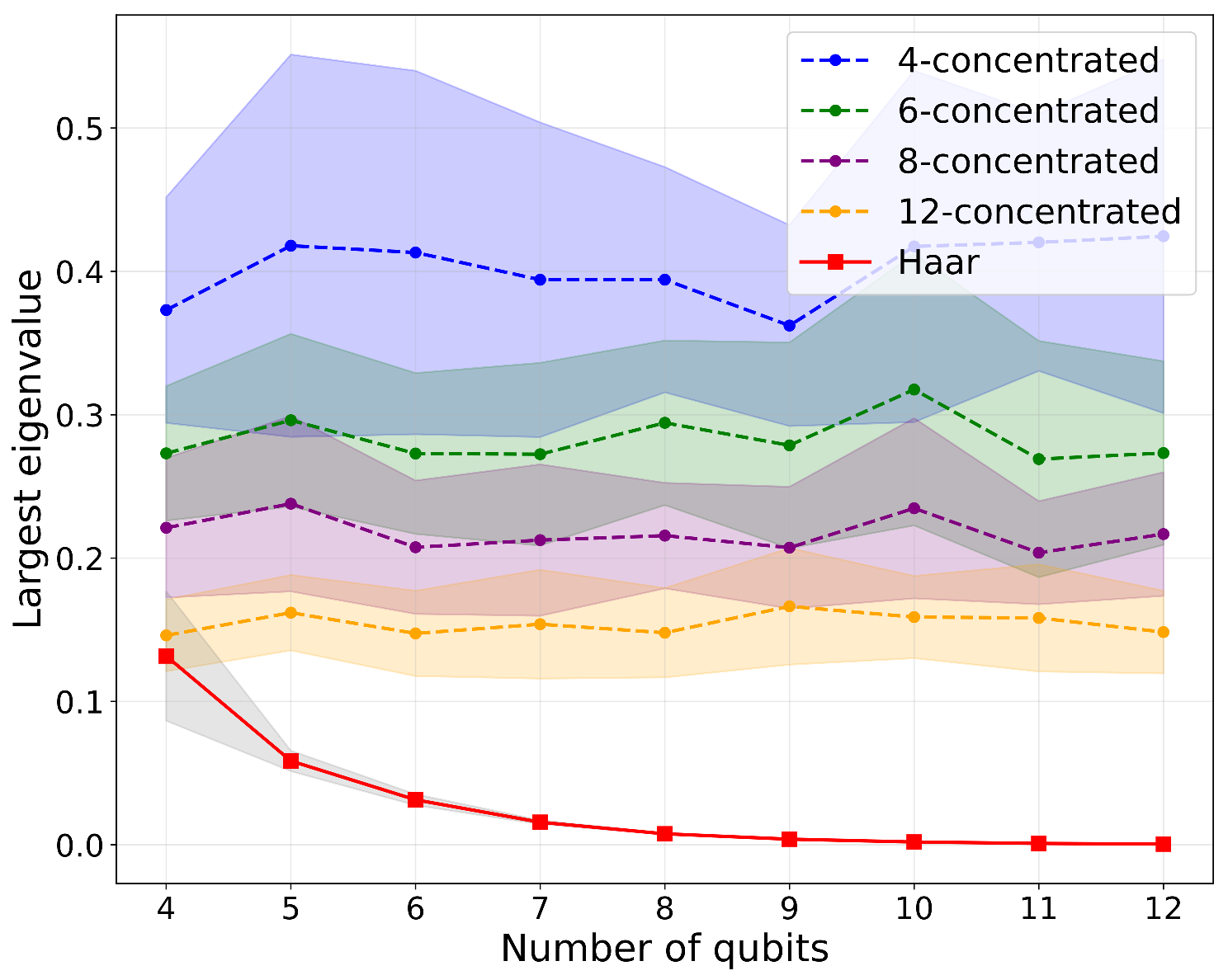}
    \caption{The scaling of the largest eigenvalues of quantum kernel integral operators with respect to the number of qubits. All data points are averaged over 30 different instances and shaded regions represent standard deviation. Concentrated positions of each concentrated unitary are chosen randomly for all instances. } 
    \label{fig:largesteigen}
\end{figure}

\begin{figure}
    \centering
    \includegraphics[width= 1.0\linewidth]{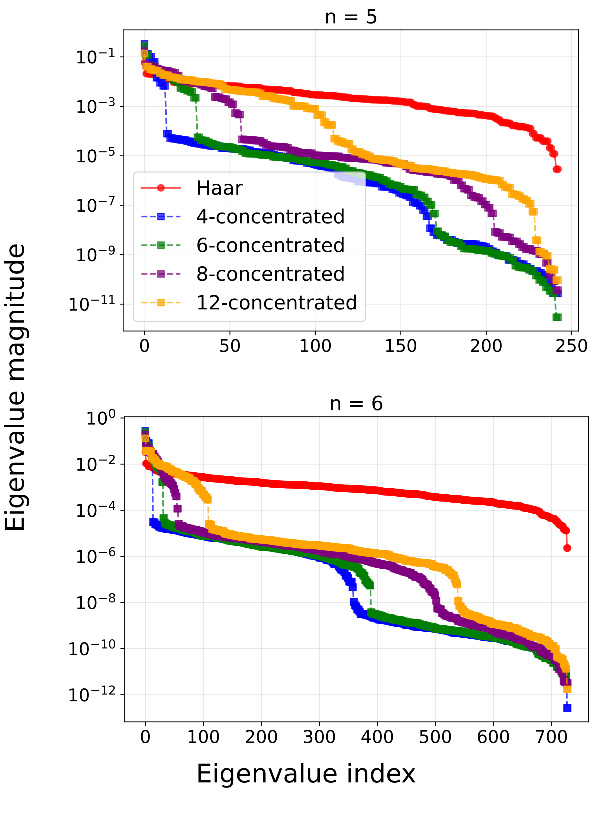}
    \caption{Eigenvalue spectra of quantum kernels satisfying two assumptions in Sec.~\ref{ss:example_onelayer}. These plots are from one instance among $30$ experiments.} 
    \label{fig:Eigenspectra}
\end{figure}

Here we study the scaling of eigenvalues of the kernel integral operators which come from given unitaries. Note that these are different from the eigenspectra of unitaries or quantum states themselves. To study, we construct 30 random kernel instances and observe the magnitudes of their largest eigenvalues as the number of qubits increases. For runs with the same sparsity $s$, the positions of the concentrated amplitudes are chosen randomly. 

Figure~\ref{fig:largesteigen} clearly displays the decay of the largest eigenvalues when the pre-encoding state comes from a Haar random unitary. Meanwhile, the largest eigenvalues from the sparse pre-encoding state does not scale with the system size, but only affected by the sparsity. In Fig.~\ref{fig:Eigenspectra} we plot all eigenvalues with a log scale. One can clearly see rather non-flat eigenspectra of the $s$-concentrated models, and their concentration to large values. The large step-like structure of spectra from concentrated models arises because only $s$ components contribute significantly to $|\psi_\alpha|^2$, and these contributions are typically much larger than those from the remaining $2^n-s$ components. 
As discussed earlier in the theoretical analysis, in principle these kernels with concentrated spectra can be approximated classically by truncating small eigenvalues, however this is only possible when the entire spectral decomposition is known, which is not typically the case.

These numerical results confirm the conclusions of our theoretical analysis that using sparse ($s$-concentrated) unitaries provides an alternative way to obtain rapid decay of quantum kernels, distinct from bandwidth-tuning strategies explored in the literature~\cite{Canatar2021spectral, canatar2023bandwidth}. Moreover, our analysis applies to more general models beyond the product-bandwidth kernels of the form  $K(x.x') = \prod_{i=1}^n  K_{i}(\lambda x_i,\lambda x'_i)$ that were previously analyzed.

\subsubsection{Generalizability comparison on synthetic dataset}
\begin{figure}
    \centering
    \includegraphics[width=\linewidth]{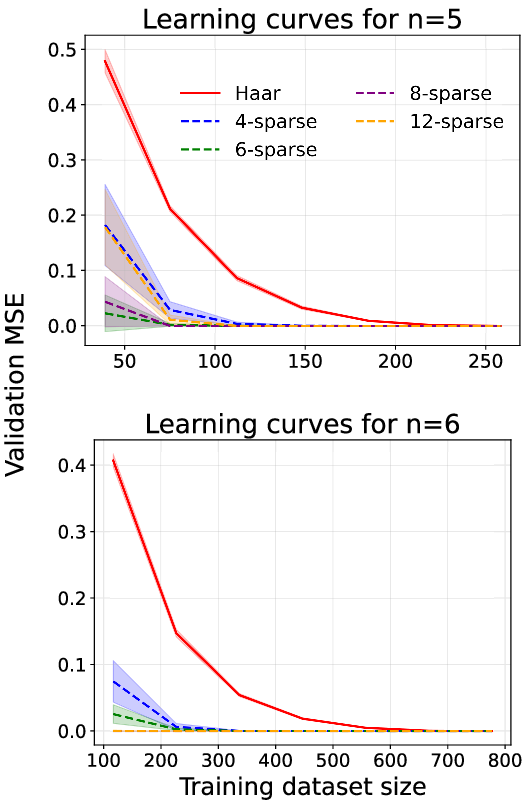}
    \caption{Learning curves for kernel ridge regression tasks on the model-tailored datasets. Data points are averaged over 30 random models, and shaded regions stand for the standard deviation. Total number of training samples is $ 2\times 3^n$, and the test set ratio was set to $0.2$. Training was performed with the KernelRidgeRegression implementation from the scikit-learn Python package. } 
    \label{fig:learning_curves}
\end{figure}

Given the ``no free lunch" theorem of ML, we cannot expect a single kernel to perform well on all learning problems. In particular, all kernels have different strengths, and a kernel can only perform well when the target function has a large overlap with its dominant eigenspaces. In other words, even if the largest eigenvalue is relatively large, the kernel can still perform poorly on some datasets if the target function is not aligned with those eigenvectors. To isolate the effect of largest-eigenvalue scaling, we construct datasets \textit{tailored} to each $s$-concentrated model and to the Haar model. First, let $P :=1+n +n(n-1)/2$ (a number of eigenvalues that $\gtrsim 1/2^{n+2}$ for typical Haar random models). For a given unitary $W_H$, and $W_s$, we select the P eigenfunctions $\{\cos(\omega^\alpha\cdot x)\}_\alpha$ associated with the $P$ largest eigenvalues of the kernel defined by the $W$ operators. We then draw a random $P$-dimensional coefficient vector $c_{\alpha,W}$ whose entries have the same magnitude order to the model's, and construct the target function $f_{t,W}(x) = \sum_{\alpha}(\sqrt{c_{t,W}})_\alpha\cos(\omega^{\alpha}\cdot x)$.  By using these model-specific target functions, the target–model alignment across experiments is comparable, so each kernel is effectively evaluated under near best-case alignment.

Figure~\ref{fig:learning_curves} shows the average validation mean squared error (MSE) during training. The results show that the concentrated models reach a lower validation error more quickly than the Haar model, as predicted by our prior theoretical analysis. Here, we once more emphasize that each model—constructed from a different unitary—works on target functions that were built to lie in that model’s top $P$-dimensional eigenspace. Therefore, these experiments do not imply the general claim that concentrated models are always superior to Haar models. For an extreme example, when the target function lies in the subspace spanned by several of the lowest eigenvectors of a concentrated model, the Haar model could trivially outperform the concentrated model. However, what it does confirm is that when the necessary criteria of kernel-target alignment are satisfied, kernels with more concentrated spectra generalize more rapidly than kernels with uniform spectra. We also discussed potential sources for notably bad performance of the four-sparse model at an early stage in Appendix~\ref{app:4-sparse}. 

Taken together, these numerical experiments confirm the design principles suggested by our theoretical analysis. Specifically, when using kernels satisfying Assumptions 1 and 2 in the previous section, in order to generalize well it is necessary (but not sufficient) to use kernels with a concentrated pre-encoding state.

\section{Summary and Discussions}\label{sec:conclusion}

In this work we have introduced the concept of an \textit{entangled tensor kernel} (ETK), a generalization of product kernels. These ETKs allow one to construct new kernels from a set of existing feature maps, by using a \textit{core matrix} to define an inner product on the tensor product of feature spaces of the original feature maps. This core matrix can be understood as creating an entangled feature map. Moreover, we have discussed the complexity of evaluating ETKs, showing that in some cases they can be efficiently evaluated via tensor networks. 

Having defined ETKs, we then showed that all quantum kernels are ETKs, whose local feature maps are defined by the data-encoding strategy of the quantum circuit. This fact then allowed us to obtain a variety of insights into quantum kernels by examining them through the lens of their ETK representations.

Our first insight in this regard is that quantum kernels can provide a mechanism for the efficient evaluation of ``super-polynomial bond-dimension'' ETKs. These are ETKs whose core tensor does not admit a representation as an MPO with polynomial bond-dimension and, therefore, cannot be straightforwardly evaluated classically in an efficient way. In this sense, one can think of quantum kernels which generate super polynomial bond-dimension ETKs as possessing an \textit{inductive bias} which is not straightforwardly accessible to efficient classical kernels with the same feature maps. This notion therefore provides a different lens through which to understand the potential advantages and applications of quantum kernels.

Complementing the above insights into the potential advantages of quantum kernels which generate super polynomial bond-dimension ETKs, we have also provided some analysis of whether there exist quantum kernels which are both super polynomial bond-dimension ETKs, and have the capacity to generalize well for specific problems from polynomially sized datasets. Indeed, in order to generalize efficiently one requires that the RKHS of the quantum kernel be effectively of only polynomially large dimension~\cite{Canatar2021spectral,canatar2023bandwidth}, which seems potentially incompatible with inefficiency of classical evaluation. We have shown that while a polynomially large effective RKHS is required, \textit{obtaining} the description of the relevant subspace of the RKHS can be computationally hard, and even if one knows the description, evaluating the eigenfunctions of the kernel can also be computationally hard. Thus, we conjecture that there exist quantum kernels that both lead to super polynomial bond-dimension ETKs, \textit{and} generalize efficiently for some class of problems.

Additionally, the fact that all quantum kernels admit an ETK representation, naturally suggests the use of ETKs as \textit{classical surrogates} for the dequantization of quantum kernel methods, and we discuss the potential and limitations of this approach. In particular, we provide a variety of concrete directions for future work in this direction.

Finally, to further showcase the power of the ETK lens for the analysis of quantum kernels, we have used the ETK framework for the analysis of a class of concrete quantum kernels which goes beyond those analyzed in previous works~\cite{Kubler2021inductive,Shaydulin_2022,canatar2023bandwidth}. In particular, we have used the ETK framework to obtain the Mercer decomposition for the relevant class of quantum kernels, from which one can gain insights into the spectrum of the kernel. This then itself informs design choices which effect the ability of the kernel to generalize well from only polynomially sized datasets.

Taken together, we see that ETKs provide a novel lens for the analysis of quantum kernels, which helps us to understand both the potential advantages and limitations of quantum kernel methods. It is our hope that the ETK framework can be of use in the further analysis and design of both classical and quantum kernels.

\acknowledgments

The authors would like to thank Elies Gil-Fuster and Jens Eisert for helpful conversations and constructive feedback. R.S. is grateful for support from the Alexander von Humboldt Foundation, under the German Research Chair program. This research was supported by the education and training program of the Quantum Information Research Support Center, funded through the National Research Foundation of Korea (NRF) by the Ministry of Science and ICT (MSIT) of the Korean government (Grants No. RS-2023-NR057243, No. RS-2024-00413957, No. RS-2024-00438415, No. RS-2023-NR076733, No. RS-2025-25464492, and No. RS-2024-00437191), and by the Institute of Information \& Communications Technology Planning \& Evaluation (IITP) grant funded by the Korea government (MSIT) (RS-2025-25464252, IITP-2025-RS-2020-II201606 and IITP-2025-RS-2024-00437191).

\section*{Data Availability}

The data that support the findings of this article are not publicly available upon publication because it is not technically feasible and/or the cost of preparing, depositing, and hosting the data would be prohibitive within the terms of this research project. The data are available from the authors upon reasonable request.

\appendix

\section{Examples of entangled tensor kernels.}\label{app:examples_of_ETK}

We supplement the definition of entangled tensor kernels in Section~\ref{ss:ETK} with a variety of concrete examples.

\subsection{Polynomial kernel}
The polynomial kernel $K_{(N,c)}$ is defined via
\begin{equation}
    K_{(N,c)}(\vx,\vx') = (c + \vx^\top\vx')^N.
\end{equation}
One can immediately see that 
\begin{equation}
K_{(N,c)}(\vx,\vx') = \prod^N K_c(\vx,\vx'),
\end{equation}
where 
\begin{equation}
    K_{c}(\vx,\vx') = c + \vx^\top\vx'.
\end{equation}
As such, the polynomial kernel $K_{(N,c)}$ is a product kernel. Note, that from the feature map perspective, we can write
\begin{equation}
K_{(N,c)}(\vx,\vx') = \langle F(\vx)|F(\vx')\rangle, 
\end{equation}
where 
\begin{equation}
\ket{F(\vx)} = \bigotimes^{N}|F_{c}(\vx)\rangle,
\end{equation}
and 
\begin{equation}
\ket{F_c(\vx)} =  [ \sqrt{c},~ \vx_1,~ \vx_2,~\ldots,~ \vx_d].
\end{equation}

\subsection{Linear sum kernels}
It is a well-known fact that any linear sum of kernels with positive coefficients is also a valid kernel, i.e.
\begin{equation}
    K_{(N,\mathbf{a})}(\vx,\vx') := \sum_{i=1}^N\mathbf{a}_iK_i(\vx,\vx')
\end{equation} is a valid kernel.

To represent this kernel as an ETK, we construct a product feature map
\begin{align}
    \ket{F(\vx)} = \bigotimes_{i=1}^N\begin{pmatrix}
        1\\F_i(\vx)
    \end{pmatrix}.
\end{align}
Additionally, let us denote the dimensions of local feature maps as $d_i$. For notational simplicity we define the multi-dimensional index $h := h_1h_2\ldots h_N$, where $h_i\in\{0,1,\ldots,d_i-1\}.$ Then,
\begin{align}
    K_{(N,\mathbf{a})}(\vx,\vx') &= \sum_{i=1}^N\mathbf{a}_i\langle F_i(\vx)|F_i(\vx')\rangle\\
    &= \bra{F(\vx)}C\ket{F(\vx')},
\end{align} where the core tensor is diagonal with
\begin{align}
    C_{h,h} = \begin{cases}
        a_i\quad\text{if}\quad h_i \neq 0, \text{and}~h_{k\neq i}=0 \\
        0\quad \text{else}
    \end{cases}.
\end{align} One can note that linear-sum kernels are a subclass of ETK where we only use first-order feature maps among all possible correlated feature maps such as $\ket{F_i(\vx)}\ket{F_j(\vx)}$, $\ket{F_i(\vx)}\ket{F_j(\vx)}\ket{F_k(\vx)}$ and so on.

\subsection{Shift-invariant kernels}

A shift invariant kernel is one satisfying
\begin{equation}
K(\vx,\vx') = K(\vx-\vx')
\end{equation}
for some positive semi-definite function $K:\mathcal{X}\mapsto\mathbb{R}$. If we consider $\mathcal{X}=[-\pi,\pi]$, then any such shift invariant kernel can be written as
\begin{equation}
K(x-x') = \sum_{j = 0}^{\infty} \gamma_j\cos(j(x-x')),
\end{equation}
with $\gamma_j\geq 0$~\cite{Wainwright2019}. Here, we show that all shift invariant kernels on $\mathcal{X}$ with a \textit{finite frequency cutoff} -- i.e. with $\gamma_j = 0$ for all $j> N$ -- can be written as entangled tensor kernels, with only $ \tilde{N}:=\lceil\log_3(2N+1)\rceil$ local feature maps. To this end, we start by defining the local feature maps,
\begin{equation}\label{eq:exponential_featuremap}
    \ket{F^{(k)}(x)} = \begin{pmatrix}
        e^{-i3^{k-1}x}\\
        1\\
        e^{i3^{k-1}x}
    \end{pmatrix},
\end{equation}
for $k\in [\tilde{N}]$. As per section~\ref{ss:ETK}, we then define 
\begin{equation}
    \ket{F(x)} = \bigotimes_{k}\ket{F^{(k)}(x)} \in \mathbb{C}^{{3}^{\tilde{N}}}.
\end{equation}
The components of $\ket{F(x)}$ can be indexed with trit-strings $h\in\{0,1,2\}^{\tilde{N}}$, and one can verify that
\begin{align}
 F_h(x) &= \exp(\left[\sum_{k}^{\tilde{N}} 3^{k-1}(h_k-1)\right]ix)\\
 &=\exp(i \alpha_h x),
\end{align}
where we have defined
\begin{equation}
\alpha_h = \sum_{k}^{\tilde{N}} 3^{k-1}(h_k-1).
\end{equation}
We note that 
\begin{equation}\begin{split}
\{\alpha_h\,|\,h\in\{0,1,2\}^{\tilde{N}}\} = \Bigg\{-\left(\frac{3^{\tilde{N}}-1}{2}\right),\ldots\\,-1,0,1,\ldots, \left(\frac{3^{\tilde{N}}-1}{2}\right)\Bigg\},
\end{split}
\end{equation}
and that $\alpha_{\overline{h}} = -\alpha_{h}$, where 
$\overline{h}$ is defined to be the trit string obtained by exchanging all occurrences of 0 with 2 and vice versa, while leaving 1s unchanged. Using this, let $C$ be the $3^{\tilde{N}}\times 3^{\tilde{N}}$ diagonal matrix defined via
\begin{equation}\label{eq:core_example}
C_{h,h} = \begin{cases} 0 \,\,\text{ if } |\alpha_h| > N \\
\gamma_j \text{ if } |\alpha_h| = j \leq N.
\end{cases}
\end{equation}
We then have
\begin{align}
\bra{F(x)}C\ket{F(x')}
& = \sum_{h\in\{0,1,2,3\}^{\tilde{N}}} C_{h,h}e^{i\alpha_h(x-x')}\\
&= \gamma_0 + \sum_{j=1}^{N}\gamma_j\left(e^{ij(x-x')} + e^{-ij(x-x')} \right)\\
&= \sum_{j=0}^{N}\gamma_j\cos(j(x-x'))\\
&=K(x-x').
\end{align}
As such, $K$ is indeed an entangled tensor kernel, with local feature maps $\{F^{(k)}\,|\,k\in\tilde{N}\}$ defined via Eq.~\eqref{eq:exponential_featuremap}, and core tensor $C$ defined via Eq.~\eqref{eq:core_example}. 

As per the discussion in Section~\ref{ss:complexity}, naively evaluating $K$ requires $N$ operations. However, if the core tensor $C$ admits a representation as an MPO with $O(\log N)$ bond-dimension, then the ETK representation can be used to evaluate $K(x,x')$ using only $O(\log N)$ operations. Additionally, we note that the ETK representation presented here is \textit{not} unique. For instance, one could use $[e^{-ix}~1~e^{ix}]^\top$ as local feature maps, but this choice would require $N$ local feature maps. We use this observation to highlight the non-uniqueness of the ETK representation of a given kernel, and the strong dependence of computational/memory complexity on the representation chosen.

\section{Obtaining Mercer decompositions of entangled tensor kernels}\label{app:mercer_decomp}

As discussed in Section~\ref{sss:inductive_bias}, one can extract a variety of quantitative insights into the inductive bias of a kernel from its \textit{Mercer decomposition}. In light of this, we provide here an algorithm for obtaining the Mercer decompositions of ETKs. We stress from the outset however that the complexity of this algorithm scales with the product of the sizes of the local feature spaces used in the ETK, and is therefore often not efficient in practice. 

To start, let us assume that we are given an ETK
\begin{equation}
K_C(\vx,\vx') = \bra{F(\vx)}C\ket{F(\vx')},
\end{equation}
with 
\begin{equation}
\ket{F(\vx)} = \bigotimes_k \ket{F^{(k)}(\vx)}.
\end{equation}
We assume $\ket{F^{(k)}(\vx)} \in\mathbb{C}^{d_k}$, so that $\ket{F(\vx)}\in\mathbb{C}^{d}$, where $d=\prod_k d_k$. As before, we denote the components of $F:\mathcal{X}\rightarrow\mathbb{C}^d$ with $F_i:\mathcal{X}\rightarrow\mathbb{C}$, for all $i\in[d]$. We note that all component functions are elements of the vector space of square integrable functions $L^2(\mathcal{X},P_\mathcal{X})$.

A first simple case is the one in which $C$ is diagonal, and all components of $F$ are orthonormal. By comparison with Eq.~\eqref{eq:mercer_form}, we see that this kernel is already in Mercer form. However, in general the core tensor $C$ will not be diagonal, and the component functions~$\{F_i\}$ will not be orthonormal. In this case, we proceed via the following steps:

\textbf{Step 1: Obtain an orthonormal basis from component functions.} We start by defining
\begin{equation}
\tilde{d} = \text{dim}\left[\text{span}(\{F_i\,|\,i\in [d]\})\right].
\end{equation}
Additionally, without loss of generality we assume an ordering such that $\{F_i\,|\, i\in[\tilde{d}]\}$ are linearly independent (one simply re-orders rows of $C$ appropriately). By using the Gram-Schmidt orthonormalization procedure, we can construct an orthonormal basis $\{e_i\,|\,i\in[\tilde{d}]\}$ for $\text{span}(\{F_i\,|\,i\in [d]\})$ by first constructing
\begin{equation}
\begin{split}
    u_i = F_i - \sum_{j=1}^{i-1}\frac{\left<u_j,F_i\right>}{\left<u_j,u_j\right>}u_j,
    \end{split}
\end{equation}
and then defining $e_i = u_i/\|u_i\|_2$, where $\|u_i\|_2 = \sqrt{\langle u_i,u_i\rangle}$. If we then define 
\begin{equation}
\ket{\tilde{e}(\vx)} = (e_1(\vx),\ldots,e_{\tilde{d}}(\vx))^\top,
\end{equation}
one has that
\begin{equation}
\ket{\tilde{e}(\vx)} = \tilde{L}\ket{F(\vx)},
\end{equation}
where $\tilde{L}$ is a $\tilde{d}\times d$ lower-triangular matrix with
\begin{equation}
\begin{split}
        \tilde{L}_{i,i}&= \frac{1}{\norm{u_i}_2}\\
        \tilde{L}_{i,i-j} &= -\sum_{l=0}^{j-1} \tilde{L}_{i,i-l}\frac{\langle u_{i-j},F_{j-l}\rangle}{\langle u_{i-j},u_{i-j}\rangle},\quad (1\leq j \leq i-1).     
\end{split}
\end{equation}
\textbf{Step 2: Pad the orthonormal basis.} Now, we note that by assumption on the dimension of $\text{span}(\{F_i\,|\,i\in [d]\})$ and the ordering of $\{F_i\}$, for all $i>\tilde{d}$ one has that 
\begin{equation}
F_i = \sum_{j=1}^{\tilde{d}}\alpha_{i,j}F_j,
\end{equation}
for some $\{\alpha_{i,j}\in\mathbb{C}\,|\,j\in[\tilde{d}]\}$.
Using this, we can construct a $d\times d$ lower triangular matrix $L$ by adding rows to $\tilde{L}$, via 
\begin{equation}
L_{i,j} = \begin{cases}
\tilde{L}_{i,j} \,\quad\,&\text{ if } i \leq \tilde{d}\\
-\alpha_{i,j} \,&\text{ if } i > \tilde{d}, j \leq{\tilde{d}}\\
1 \quad\,\quad\,&\text{ if } i=j >\tilde{d}.
\end{cases}
\end{equation}
This matrix is lower triangular, with non-zero entries on all diagonals, and is therefore invertible (which, as we will see in Step 3, was the motivation for the padding). Additionally. we have that
\begin{align}
L\ket{F(\vx)} &= \begin{pmatrix}e_1(\vx)\\
\vdots\\
e_{\tilde{d}}(\vx)\\
0\\
\vdots\\
0
\end{pmatrix},\\
&:=\ket{e(\vx)}
\end{align}

\textbf{Step 3: Transform and truncate $C$.} Using the above, we have that
\begin{align}
K_C(\vx,\vx) &= \bra{F(\vx)}C\ket{F(\vx)} \\
&=\bra{F(\vx)}L^\top (L^{\top})^{-1}CL^{-1}L\ket{F(\vx)}\\
&=\bra{e(\vx)}(L^{\top})^{-1}CL^{-1}\ket{e(\vx)}\\
&= \bra{e(\vx)}\hat{C}\ket{e(\vx)}\\
&= \bra{\tilde{e}(\vx)}\tilde{C}\ket{\tilde{e}(\vx)},\label{eq:almost_mercer}
\end{align}
where $\hat{C} = (L^{\top})^{-1}CL^{-1}$, and $\tilde{C}$ is the upper left $\tilde{d}\times\tilde{d}$ block of $\hat{C}$.

\textbf{Step 4: Diagonalize $\tilde{C}$.} The expression in Eq.~\eqref{eq:almost_mercer} is almost in Mercer form. The final step is to diagonalize $\tilde{C}$. More specifically, to find $U$ such that $\tilde{C} = U^{\dagger}DU$. Given this, we then have that
\begin{align}
K_C(\vx,\vx) &= \bra{\tilde{e}(\vx)}\tilde{C}\ket{\tilde{e}(\vx)} \\&= \bra{\tilde{e}(\vx)}U^{\dagger}DU\ket{\tilde{e}(\vx)}\\
&=\bra{\hat{e}(\vx)}D\ket{\hat{e}(\vx)},\label{eq:final_mercer}
\end{align}
where $\ket{\hat{e}(\vx)} := U\ket{\tilde{e}(\vx)}$. As the components of $\ket{\tilde{e}(\vx)}$ are orthonormal, so too are the components of $\ket{\hat{e}(\vx)}$. As a result, Eq.~\eqref{eq:final_mercer} is finally the desired Mercer decomposition of~$K_C$.

\section{A standard form for data-dependent quantum circuits}\label{app:decomposition_into_standard_form}

In this section we show how any data-dependent unitary can be transformed into a data-dependent quantum circuit of a ``standard form''. To make this precise, we formalize a data-dependent unitary as a map $U:\mathcal{X}\rightarrow \mathrm{SU}(2^n)$. With this convention $U(\vx)\in\mathrm{SU}(2^n)$ represents the specific unitary corresponding to $\vx\in\mathcal{X}$. At this stage, we make no assumptions on the form of this map. We now show that for any such map $U$, there exists a parameterized quantum circuit $C:\mathcal{X}\rightarrow\mathrm{SU}(2^n)$, such that $U(\vx) = C(\vx)$ for all $x\in\mathcal{X}$, and $C$ is in the \textit{standard form}
\begin{equation}\label{eq:s_form_app}
C(\cdot)=\prod_{j=1}^LS_j(\cdot)W_j
\end{equation}
where for all $j\in L$, $W_j\in \mathrm{SU}(2^n)$ is a fixed unitary, and $S_j:\mathcal{X}\rightarrow\mathrm{SU}(2^n)$ is a data-dependent unitary of the form
\begin{align}
S_j(\cdot) &:=\left(\bigotimes_{k=1}^{n}e^{-i\phi_{jk}(\cdot)Z_{k}/2}\right),\\
&:= \left(\bigotimes_{k=1}^{n}R_Z(\phi_{jk}(\cdot))\right)
\end{align} 
where $\phi_{jk}:\mathcal{X}\rightarrow[0,2\pi)$ are functions which depend on $U$.

To show this, we consider decomposing $U$ into a sequence of Givens rotations, as per Ref.~\cite{Juha2004efficient}. If one does this,  in the case where $U$ is data-dependent, one will obtain
\begin{equation}
U(\cdot) = \prod_{j = 1}^{\tilde{L}}G_j(\cdot),
\end{equation}
where for all $j$, the parameterized Givens rotation $G_j:\mathcal{X}\rightarrow\mathrm{SU}(2^n)$ is a parameterized rotation in a fixed (non data-dependent) two-dimensional subspace of the $2^n$ dimensional Hilbert space. Given this, we then note the following:

\begin{enumerate}
\item Any such rotation can be written as a multi-controlled parameterized sinqle-qubit rotation, which in turn can be decomposed into a circuit consisting only of $\mathrm{CNOT}$ gates and parameterized single-qubit gates $R:\mathcal{X}\rightarrow\mathrm{SU}(2)$. We stress that as the subspace in which the rotation occurs is fixed, the order and position of the $\mathrm{CNOT}$ gates and single qubit gates are also fixed -- i.e. they depend only on $j$, and not on~$\vx$. 
\item Each parameterized single-qubit rotation can be written as 
\begin{equation}
R(\cdot) = R_Z(\phi_{1}(\cdot))R_Y(\phi_{2}(\cdot))R_Z(\phi_{3}(\cdot))
\end{equation}
where $\phi_{i}:\mathcal{X}\rightarrow[0,2\pi)$ are functions that depend on $U$. 
\item Each $R_Y$ gate can be decomposed into $R_Z$ gates and basis rotations using the relation $Y = F^{\dag}ZF$, where $F := \frac{1}{\sqrt{2}}\begin{pmatrix}
        1 &1
        \\i &-i
    \end{pmatrix}$
\end{enumerate}
If one does all of the above, and then groups together all data-dependent gates and data-independent gates into separate layers, one finally arrives at a circuit in the form of Eq.~\eqref{eq:s_form_app}.

At this stage we stress that in the worst case, the decomposition above will lead to circuits in which 
\begin{enumerate}
\item $L$ is exponentially large in the number of qubits.
\item The functions $\{\phi_{jk}\}$ are extremely complicated.
\end{enumerate}
However, in practice, one normally starts not from some arbitrarily defined data-dependent unitary $U:\mathcal{X}\rightarrow\mathrm{SU}(2^n)$, but rather from some parameterized quantum circuit $C:\mathcal{X}\rightarrow\mathrm{SU}(2^n)$ with a polynomial number of gates, of which the data-dependent gates are all of some fixed simple form (eg: gates that encode data by parameterizing the time evolution of a simple few-body Hamiltonian~\cite{Havlicek2019,Schuld_2021}). In many such cases, the decomposition of the circuit into the standard form of Eq.~\eqref{eq:s_form_app} can be done much more efficiently than via the Givens rotation construction given above, which is simply used to demonstrate the universality of the standard form.

\section{Positive semidefiniteness and Hermiticity of $C_T$}

To ensure a tensor network representation of a quantum kernel is ETK, we need to check that the core matrix that comes from the quantum circuit contraction is positive semidefinite and Hermitian (symmetric). 

First, we observe that the operator $O'$ is a tensor
product of identity and choi matrices of unitaries that are applied in the circuit, so it is positive semidefinite (PSD) and Hermitian. The operator $\rho^\top$ is also PSD as it is a tensor product of (un-normalized) density matrices. By the Schur product theorem, $O'\odot \rho^\top=A$ is also PSD. Transpose preserves positivity and eigenvalues of the vertical tensor product of two matrices are given by multiplication between the eigenvalues of two matrices. Also, Schur product of two Hermitian (symmetric) matrices is Hermitian (symmetric), and from Eq.~\eqref{eq:vertical tensor}, we see that the vertical tensor product of hermitian matrix and its transpose preserves the Hermiticity. Therefore, $C$ is PSD and Hermitian.

Next, we show that $C_T$ is PSD and Hermitian as well. For all $\ket{v} \in \mathbb{C}^{3^N}$, and PSD $C$,
\begin{equation}
\begin{split}
    \frac{1}{4^N}\bra{v}C_T\ket{v} &= 2^N\bra{\tilde{v}}\Pi_3\Pi_{3}P_U^\dag CP_U\Pi_{3}\Pi_3\ket{\tilde{v}}\\
    &=2^N\bra{\tilde{v}}\Pi_{3}C'\Pi_{3}\ket{\tilde{v}} \geq0,
\end{split}
\end{equation}where $\Pi_3 := \begin{pmatrix}
    1&0&0&0\\
    0&1&0&0\\
    0&0&1&0\\
    0&0&0&0
\end{pmatrix}^{\otimes N}$ is a projection and $P_U = \frac{1}{\sqrt{2}}\begin{pmatrix}
         1 &0 &0&1 \\
         0 &1 &i&0\\
         0 &1 &-i&0\\
         1 &0 &0&-1
     \end{pmatrix}^{\otimes N}$, which is a unitary extension of $P$ and $\ket{\Tilde{v}}$ is an arbitrary extension of $\ket{v}$ to $\mathbb{C}^{4^n}$, which embed arbitrary elements to the extended dimension. Conjugation with unitary preserves PSD property, and $\Pi_3\ket{\Tilde{v}} \in \mathbb{C}^{4^n}$, the last inequality follows. If $H$ is Hermitian, then both $U^\dagger H U$ and $\Pi H\Pi$ are Hermitian whenever $U$ is unitary and $\Pi$ is a projection operator.

\section{Possible sources for bad performance of four-sparse model in our numerics}\label{app:4-sparse}

\begin{figure}
    \centering
    \includegraphics[width=1.0\columnwidth]{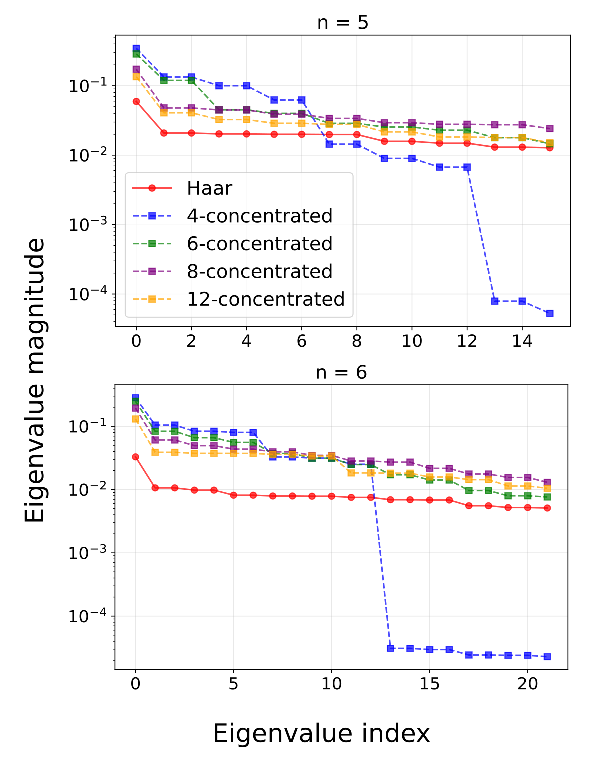}
    \caption{Eigenspectra of the largest $P$ eigenvalues of kernel models, where $P$ is the number of support of target function. } 
    \label{fig:eigenspectra_appendix}
\end{figure}

The performance of quantum kernels depends not only on the scaling of the largest eigenvalue but also on kernel–target alignment, which quantifies how well the function space spanned by the eigenfunctions with the largest eigenvalues covers the target function. In Fig.~\ref{fig:eigenspectra_appendix}, we plot the largest $P = 1 + n + n(n-1)/2$ eigenvalues for the models from one of our experiments, where $P$ is the number of basis functions of the target function. One can see sharp drops in the eigenvalues for the four-sparse model due to its high sparsity. This possibly makes it difficult for the model to learn components associated with the low-eigenvalue basis functions.

\bibliographystyle{apsrev4-2}
\bibliography{biblio}

\end{document}